\documentclass[fleqn,usenatbib]{mnras}

\usepackage{newtxtext,newtxmath}

\usepackage[T1]{fontenc}
\usepackage{enumitem}
\usepackage{ulem}

\DeclareRobustCommand{\VAN}[3]{#2}
\let\VANthebibliography\thebibliography
\def\thebibliography{\DeclareRobustCommand{\VAN}[3]{##3}\VANthebibliography}

\usepackage{graphicx}	
\usepackage{amsmath}
\usepackage{subcaption}

\usepackage{soul}
\usepackage{dirtytalk}

\title[Spin-modulated circular polarisation]{Detection of spin-modulated circular polarisation and radial velocity variations in the long-period Intermediate Polar 1RXS J080114.6–462324\thanks{Based on observations made with the Southern African Large Telescope (SALT) under program 2024-2-SCI-038 (PI: Z. N. Khangale)} }

\author[V. Moloi et al.]{
V. Moloi,$^{1,2}$\thanks{E-mail: victor@saao.ac.za (VM)}
S. B. Potter,$^{2,3}$
Z. N. Khangale,$^{1}$
D. A. H. Buckley$^{1,2,4}$
L. Booi$^{1,2}$
and P. A Woudt,$^{1}$
\\
$^{1}$Department of Astronomy, University of Cape Town, Private Bag X3, Rondebosch 7701, South Africa\\
$^{2}$South African Astronomical Observatory, PO Box 9, Observatory 7935, Cape Town, South Africa\\
$^{3}$Department of Physics, University of Johannesburg, PO Box 524, Auckland Park 2006, South Africa\\
$^{4} $Department of Physics, University of the Free State, PO Box 339, Bloemfontein 9300, South Africa\\
}

\date{Accepted 2025 October 11. Received 2025 October 08; in original form 2025 September 01}

\pubyear{\the\year{}}

\begin{document}
\label{firstpage}
\pagerange{\pageref{firstpage}--\pageref{lastpage}}
\maketitle

\begin{abstract}
We present a comprehensive photometric, spectroscopic, and polarimetric study of the intermediate polar (IP) 1RXS J080114.6–462324, using observations from the South African Astronomical Observatory (SAAO) 1.0‑m and 1.9‑m telescopes and the Southern African Large Telescope (SALT), complemented by archival TESS photometry. Photometric and photo‑polarimetric data reveal a coherent modulation at the white dwarf (WD) spin period. TESS confirms periodicities of 1307.517 s (spin) and 11.803 h (orbital).
Photopolarimetric and circular spectropolarimetric measurements show circular polarisation reaching $\sim+5$\%, modulated with the WD spin period, consistent with cyclotron emission from an accreting magnetic pole. Time‑resolved optical spectra display prominent Balmer (H$\gamma$, H$\beta$, and H$\alpha$) and He\textsc{ii} $\lambda$4686 emission features and additional He\textsc{i} emissions, all exhibiting minimal radial-velocity variations. We detect red‑shifted absorption dips adjacent to the He\textsc{ii} and H$\beta$ lines, modulated at the WD spin period; periodogram analysis of the emission lines also yields spin modulation. 
These observations indicate that the system is a disc‑fed, low‑inclination IP. The combination of circular polarisation and spin‑modulated absorption by infalling accretion curtain material supports this classification. Its comparatively long orbital period among IPs and the detection of polarised emission render 1RXS J080114.6–462324 an appealing candidate for evolutionary studies, potentially offering insight into how magnetic accretion systems evolve toward synchronism at longer orbital periods.

\end{abstract}




\section{Introduction}\label{sec:review}

Cataclysmic variables (CVs) are a class of semi-detached compact binary systems composed of a white dwarf (WD) primary and a late-type main-sequence secondary star, typically a red dwarf, often referred to as the donor star. 
In these systems, mass transfer occurs from the donor star to the WD via Roche-lobe overflow, leading to the formation of an accretion disc around the WD \citep{1995cvs..book.....W}. CVs can be broadly categorised into two subclasses, non-magnetic and magnetic CVs, and this is what determines how the material will be accreted to the WD. For non-magnetic CVs, where the magnetic field strength  $B\leq 0.1 \rm \; MG$, material from the Roche lobe filling donor star flows through the inner Lagrangian point and forms an accretion disc around the WD, as the weak magnetic field cannot disrupt the stream. 
Viscous processes within the disc transport angular momentum outward, allowing matter to spiral inward and settle onto the WD's surface \citep[see,][]{2001cvs..book.....H}.

\medskip
\noindent
In magnetic CVs, the accretion process is influenced by the magnetic field strength of the WD and result in either a partial or complete disruption of the accretion disc. There are generally two subclasses of magnetic CVs, namely Intermediate Polars (also referred to as DQ Her) and Polars (also referred to as AM Her) \citep[see,][]{1990SSRv...54..195C, 1995ASPC...85....3W}.

\medskip
\noindent
Intermediate polars (IPs) host a WD with a moderately strong magnetic field. For systems where the field strength has been measured or inferred, values typically fall in the range of $ \sim1-10 \; \rm MG$ \citep[e.g.][]{1986LNP...266...97W,1992ASPC...29..242W,1994PASP..106..209P,2006Obs...126...61S,2024MNRAS.531L..82P}. However, the true distribution of magnetic fields in IPs may be broader, including both weaker and stronger systems. For example, some IPs have an estimated surface field strengths outside this range, such as EX Hydrae ($ B \sim 0.35\;\rm MG$; \citealt{2024A&A...686A.304B}) and V405 Aurigae ($ B \gtrsim 30 \; \rm MG$; \citealt{2008ApJ...684..558P, 2019cwdb.confE..50D}). Observational limitations restrict direct field measurements, so the commonly cited $1-10 \; \rm MG$ range largely reflects which systems have detectable signatures, such as cyclotron lines, spin equilibria, or polarisation, rather than representing a strict physical boundary.

\medskip
\noindent
In these systems, the magnetic field is sufficiently strong to truncate the inner regions of the accretion disc at the magnetospheric radius, if such a disc is present. However, there exists at least one confirmed discless intermediate polar, V2400 Oph (RX J1712.6-2414) \citep[see][]{1995MNRAS.275.1028B,2002MNRAS.331..407H}. Within this radius, the ram pressure of the accretion flow becomes comparable to the magnetic pressure, forcing the inflowing material to couple onto the WD’s magnetic field lines. The gas is then funneled along these field lines in so-called accretion curtains to the magnetic poles of the WD. This produces localised accretion regions, where the material is shock-heated, generating hard X-ray and optical/UV cyclotron emission. The photometric and spectroscopic variability of IPs is generally dominated by two fundamental frequencies, the WD spin frequency and the binary orbital frequency. In addition, sideband modulations arising from beat frequencies are sometimes observed \citep[see,][]{1994PASP..106..209P}, with a few specific cases reported in the literature \citep[e.g.][]{1996MNRAS.280..937N,2002A&A...384..195N}. In almost all confirmed IPs, the WD spin period is markedly shorter than the orbital period, reflecting the asynchronous nature of accretion in these systems \citep[see eg. ][]{2004ApJ...614..349N}.  In addition, IPs display evidence of more energetic phenomena, including sudden and dramatic increases in brightness known as outbursts. These can be divided into the relatively infrequent \textit{normal} outbursts, which may last for several days (e.g., EX Hya), and the much shorter timescale events now termed \textit{micronovae}. The latter were first identified and named by \citet{2022MNRAS.514L..11S,2022Natur.604..447S}, who showed that they arise from thermonuclear runaways confined to the magnetic poles of the WD. Subsequent studies, such as \citet{2025MNRAS.539.2424V}, extended this classification to include DW Cnc and highlighted TV Col as an additional example. Such events represent some of the most energetic phases in the evolution of CVs, during which the system’s luminosity can increase substantially over several days before returning to a quiescent state \citep[see][]{2001cvs..book.....H, 2006ConPh..47..363S}.

\medskip
\noindent
Polars are characterised by a WD possessing a very strong magnetic field, typically in the range 10–230 MG \citep[e.g. see,][]{1981ApJ...243L.157S, 1985ASSL..113..151L, 1990SSRv...54..195C, 2001cvs..book.....H}. Such a strong magnetic field is sufficient to completely inhibit the formation of an accretion disc \citep{1994PASP..106..209P}. Instead, material transferred from the companion star couples directly onto the WD’s magnetic field lines and is channelled toward the magnetic poles. The high field strength also enforces synchronous rotation, locking the WD spin period to the binary orbital period. However, there are polars in which the WD spin is slightly out of synchronism. Specifically, there are systems in which the spin and orbital periods differ by less than 2\% e.g. V1500 Cyg \citep{1988ApJ...332..282S,2018MNRAS.479..341P} and CD Ind \citep{1997A&A...326..195S,2000MNRAS.316..225R}, or where the two periods differ by more than 3\% e.g. IGR J19552+0044 \citep{2017A&A...608A..36T} and Paloma \citep{2007A&A...473..511S}.

\medskip
\noindent
The spectroscopic analysis of CVs generally shows the emission of the hydrogen Balmer series, such as H$\alpha$, H$\beta$, H$\gamma$, and H$\delta$ as a result of hot gas in the accretion disc or accretion columns, due to gravitational energy of the infalling material being converted into thermal energy \citep{2017AJ....153..144O, 2020AJ....159..114O}.
Most of the observed emission lines correspond to the hydrogen Balmer series, as the accreting material is predominantly hydrogen. These lines are produced when the gas is heated in the accretion disc and accretion columns. Their magnetic nature is revealed through their emission of high-ionisation lines such as He\,\textsc{ii} and C\,\textsc{iii}/N\,\textsc{iii} blend at 4650 \AA{} \citep[see][]{1995cvs..book.....W,2023MNRAS.521.2729R}. IPs are known to be characterised by the equivalent width H$\beta[EW] > 20$ \AA{} and He\,\textsc{ii} 4686 \AA{}/H$\beta> 0.4$ \citep[eg., see][]{1995cvs..book.....W,2006A&A...459...21M,2010A&A...519A..96M}. 

\medskip
\noindent
Further classification of magnetic CVs in terms of their magnetic nature and strength comes from two kinds of observational techniques, which include the detection of any polarised emission or the Zeeman splitting of the spectral lines. This is a direct indication of a highly magnetised WD. As shown in \citet{2012A&A...546A.104T}, the magnetic properties of CVs, such as the magnetic field strength for highly polarised systems such as polars, can then be obtained from the actual spectrum of the emitted cyclotron emission. 
The spacing between two adjacent cyclotron humps can directly give us the magnetic strength in the accretion region. In contrast to Polars, IPs have a relatively lower magnetic field strength ($\lesssim$ 10\,MG), so these features are not seen; as a result, other indirect measures are used to infer the magnetic field strength, for example the detection of circular, and sometimes linear, polarisation \citep[see e.g.][]{2009A&A...496..891B,2010ApJ...724..165K}. 

\subsection{Previous Observation of 1RXS J080114.6–462324}

1RXS J080114.6–462324, also known as PBC J0801.2–4625 was initially detected as an X-ray source in data from the INTEGRAL/IBIS (IBIS; \citealt{2003A&A...411L.223G}) survey and subsequently observed with the IBIS instrument, where it was recognized as a variable object \citep{2010MNRAS.403..945L}. 
Further optical follow-up and spectroscopic analysis led to its classification as a CV, specifically a dwarf nova\footnote{Dwarf novae are a subclass of CVs in which a  WD accretes matter via an accretion disc from a companion star. They exhibit recurrent outbursts caused by thermal-viscous instabilities in the disc, leading to sudden increases in brightness of 2–6 magnitudes.} in the optical study of the INTEGRAL objects presented in \citet{2010A&A...519A..96M}. This classification was based on the distinct features observed in its optical spectrum, such as strong emission lines typically associated with accretion processes in binary systems, including the hydrogen Balmer emission series such as H$\alpha$, H$\beta$, and H$\gamma$. 
Their magnetic nature is seen through their emission of high-ionisation lines of  He\,\textsc{ii} 4686 \AA{} and C\,\textsc{iii}/N\,\textsc{iii} blend at 4650 \AA. According to \citet{2010A&A...519A..96M}, these spectral characteristics were key in identifying the nature of the source. \citet{2017MNRAS.470.4815B} reported this system to have a WD spin period of $1310.9 \pm1.5$ s in the X-ray band and $1306.3\pm0.9$ s in the V band with no indication of the binary orbit, and strongly suggested this to be an IP CV. Through optical observations, \citet{2018AJ....155..247H} provided a refined WD spin period of $1307.55 \pm 0.10$ s using data obtained with the SAAO 1.0-m telescope. 
Further TESS observations presented by  \citet{2024MNRAS.530.3974I} revealed a periodic signal at $5.906\pm0.003$ h, which was suggested as the orbital period of the binary. The authors further reported that this system undergoes what is suspected to be a micronova outburst using the study of the simultaneous observations from ASAS-SN\footnote{The All-Sky Automated Survey for Supernovae (ASAS-SN; \citealt{2014ApJ...788...48S,2017PASP..129j4502K}) is a network of ground-based optical telescopes designed to observe the entire visible sky.} and TESS, \citep[see][for more information]{2024MNRAS.530.3974I}.

\section{Observations and Data Reduction}

A detailed log of all of the observations is provided in Table \ref{tab:Photometric_table}.

\begin{table*}
\centering
\caption{Log of all observations for 1RXS J080114.6–462324, including spectroscopic, photometric, photopolarimetric, and circular spectropolarimetric data.}
\label{tab:Photometric_table}
\begin{tabular}{lcccccc}
\hline
\hline
\textbf{Date}                          & \textbf{Telescope}         & \textbf{Instrument}     & \textbf{Filter \textbackslash Bandpass}        & \textbf{Exposure (s)}            & \textbf{Total time (hrs)}                 \\
\hline
\hline
2017/03/29                            & 1.9-m SAAO         & SPUPNIC            & 4200-5400\AA               &       200                &         4.60                         \\
2017/03/30                            & 1.9-m SAAO         & SPUPNIC            & 4200-5400\AA               &       200                &         3.90                         \\
2017/03/31                            & 1.9-m SAAO         & SPUPNIC            & 4200-5400\AA               &       200                &         4.40                         \\
2017/04/01                            & 1.9-m SAAO         & SPUPNIC            & 4200-5400\AA               &       200                &         5.00                         \\
2017/04/02                            & 1.9-m SAAO         & SPUPNIC            & 4200-5400\AA               &       200                &         5.00                         \\
2017/04/03                            & 1.9-m SAAO         & SPUPNIC            & 4200-5400\AA               &       200                &         5.50                         \\
2018/02/24                            & 1.0-m SAAO         & SHOC               & unfiltered                 &       1                  &         1.64                         \\
2019/01/04                            & 1.9-m SAAO         & HIPPO              & clear                         &        -               &         5.60                         \\
2019/01/07                            & 1.9-m SAAO         & HIPPO              & clear                        &        -                 &         5.60                         \\
2019/01/06                            & 1.0-m SAAO         & SHOC               & unfiltered                 &       10                 &         2.85                          \\
2019/01/07                            & 1.9-m SAAO         & HIPPO              & OG570 (broad red filter)    &        -                 &         6.70                         \\
2019/02/02 - 2019/02/27               & TESS              & TESS Photometer    & 6000-10000\AA              &        120               &         591.30                       \\
2021/05/11                            & 1.9-m SAAO         & HIPPO              & clear filter               &        -                 &         3.40                         \\
2023/01/18 - 2023/02/12               & TESS              & TESS Photometer    & 6000-10000\AA              &        120               &         610.20                       \\
2023/02/12 - 2023/03/10               & TESS              & TESS Photometer    & 6000-10000\AA              &        120               &         617.10                       \\
2025/01/21                            & SALT              & RSS                & PC03850                    &       300                &         0.58                            \\
2025/01/29                            & 1.9-m SAAO         & SPUPNIC            & 6200-7300\AA               &       200                &         7.55                         \\
2025/01/30                            & 1.9-m SAAO         & SHOCNWONDER        & Clear                      &        2                 &         7.12                         \\
2025/01/31                            & 1.9-m SAAO         & SHOCNWONDER        & r-filter                   &        10                &         7.63                         \\
2025/02/01                            & 1.9-m SAAO         & SHOCNWONDER        & g-filter                   &        10                &         8.14                         \\
2025/02/02                            & 1.9-m SAAO         & SPUPNIC            & 4200-5400\AA               &       200                &         7.73                         \\
2025/01/14 - 2025/02/11               & TESS              & TESS Photometer    & 6000-10000\AA              &       120                &         666.54                       \\
2025/02/03                            & SALT              & RSS                & PC03850                    &       300                &         0.58                            \\
2025/02/11 - 2025/03/12               & TESS              & TESS Photometer    & 6000-10000 \AA             &        120               &         691.36                       \\
\hline
\hline
\end{tabular}
\end{table*}

\subsection{Time-series Photometry}

\subsubsection{SAAO 1.0-m and 1.9-m Telescopes}

High-cadence (1-20s) photometric observations of 1RXS J080114.6–462324 were carried out using the Sutherland High-Speed Optical Camera (SHOC; \citealt{2013PASP..125..976C}) and SHOCnWonder\footnote{For more information on this new SHOC instrument, see: \url{https://topswiki.saao.ac.za/index.php/SHOCnWonder}} mounted on the SAAO 1.0-m and 1.9-m telescopes, respectively. The 1.0-m observations were unfiltered and conducted in 2018 and 2019. The 1.9-m observations included clear, r, and g filters and were conducted in 2025, coincident with observations made with TESS (see Table \ref{tab:Photometric_table} for details and Fig. \ref{fig:1m_1.9m_tess}). 
Calibrated aperture photometry was performed using the IO\_Phot pipeline\footnote{IO\_Phot is a Python-based pipeline for aperture photometry, developed at SAAO and is available at \url{https://www.saao.ac.za/~sbp/IO_phot/IO_phot.py}.}. IO\_Phot is a Python-based data reduction pipeline specifically designed to handle observations obtained with the SAAO SHOC cameras. All times were corrected for the light-travel time to the barycentre of the Solar System \citep[eg., see][]{2010PASP..122..935E}.

\subsubsection{TESS}

Further photometric observations were obtained from the Transiting Exoplanet Survey Satellite (TESS; \citealt{2015JATIS...1a4003R}), utilising the TESS photometer with a cadence of approximately 120 seconds. Observations span five sectors: Sector 8 (2019 February 2 - 2019 February 27), Sector 61 (2023 January 18 – 2023 February 12), Sector 62 (2023 February 12 – 2023 March 10), Sector 88 (2025 January 14 – 2025 February 11), and Sector 89 (2025 February 11 – 2025 March 12). These data are publicly accessible via the Mikulski Archive for Space Telescopes (MAST; \citealt{2015AAS...22533659F}). MAST provides two types of TESS generated light curves; Simple Aperture Photometry (SAP; \citealt{2012PASP..124..963K}), as well as PDCSAP which applies corrections to the SAP light curves for instrumental systematics, removes isolated outliers, and further adjusts the source flux for crowding effects, \cite[see][for further details]{2024ApJ...966..158P}. For this study, we used the PDCSAP TESS light curves with a quality flag of zero, to avoid non-real detections or systematic errors. The TESS data were converted from counts per second to apparent magnitudes using the formula provided in the \textit{TESS Instrument Handbook}:
\begin{equation}
    m = -2.5 \log_{10}(\mathrm{counts\,s^{-1}}) + 20.44,
\end{equation}
where the zeropoint of 20.44\, mag has an associated uncertainty of 0.05\, mag, and this is further detailed in \cite[see][for further details]{2023ApJ...956..108F}. The error measurements in magnitudes were derived through error propagation.

\subsection{Spectroscopy}

We obtained two sets of spectroscopic observations for our target 1RXS J080114.6–462324. These observations were conducted using the SPUPNIC spectrograph \citep{2019JATIS...5b4007C}, which is mounted on the SAAO 1.9-m telescope. 
The first set was taken between 2017 March 29 and April 3, using grating 4 on all six nights, which has a dispersion of 0.62 (\AA/pixel). These observations covered a wavelength range of 4200--5400\,\AA, with a slit width of 1.80$^{''}$ for all the exposures. Each exposure for our target was 200 seconds, with a 10-second exposure for the Copper-Argon (CuAr) comparison lamp. Additionally, a standard star, LTT 3218, was also observed during the same observing run. The second set was obtained on 2025 January 29 and 2025 February 2, employing grating 5 (dispersion of 0.53 (\AA/pixel)) on the first night and grating 4  on the second. The observations from the 29 January were covering the wavelength range 6200-7300\AA, with a slit width of 1.65$^{''}$. Each exposure for our target was 200 seconds, with a 10-second exposure for the Copper-Neon (CuNe) comparison lamp (see Table \ref{tab:Photometric_table}). All times were corrected for the light-travel time to the barycentre of the Solar System \citep[eg., see][]{2010PASP..122..935E}. 

Data reduction was carried out using standard procedures within the IRAF software package\footnote{IRAF (Image Reduction and Analysis Facility) is distributed by the National Optical Astronomy Observatory and is publicly available at \url{https://iraf-community.github.io/}.}, for more details \citep[see][]{1986SPIE..627..733T}.  The resulting time-averaged spectra obtained using grating 4 and 5 are presented in Fig. \ref{fig:average_spectrum}, and Fig. \ref{fig:H_alpha} respectively, and will be discussed in Section \ref{sec:spectroscopy_discussion}. 

\subsection{Polarimetry}

Polarimetric observations were obtained with the HI-speed Photo-POlarimeter (HIPPO; \citealt{2010MNRAS.402.1161P}), a high-speed photopolarimeter mounted on the SAAO 1.9-m telescope, designed to perform simultaneous photometry and polarimetry. The instrument was operated in its all-Stokes mode, enabling concurrent measurements of linear and circular polarisation, together with photometry. Three observing runs are presented in this study: the first, on 2019 January 4, using a clear filter (3500–9000 \AA); the second, on 2019 January 7, using a broad red band (5700–9000 \AA) OG570 filter; and the third, on 2021 May 11, using a clear filter. All observations were conducted with an integration time of 1 ms, providing high time-resolution polarimetric measurements. Data reduction was carried out using dedicated software written in the C programming language, as described in detail by \citet{2010MNRAS.402.1161P}. Photometric calibrations were not performed; therefore, the fluxes are presented in total counts. All times were corrected for the light-travel time to the barycentre of the Solar System \citep[eg., see][]{2010PASP..122..935E}. The resulting photometric and polarimetric light curves from the 2019 January 4 observations are presented in Fig. \ref{fig:Photopolarimetric} and will be discussed in Section \ref{sec:photopol_discussion}.

\subsection{SALT circular spectropolarimetry}

Spectropolarimetric data were obtained with the Southern African Large Telescope (SALT; \citealt{2006SPIE.6267E..0ZB}) on 2025 February 3  using the Robert Stobie Spectrograph (RSS; \citealt{2003SPIE.4841.1463B,2003SPIE.4841.1634K}) in circular spectropolarimetry mode \citep{2003SPIE.4843..170N}. A total of six exposures of 300 s each, containing the ordinary (O) and extraordinary (E) beams, were obtained, covering slightly over one spin cycle of the target. The PG0900 grating and a grating angle of 14.5$^{\circ}$, giving a spectral resolution of $\sim$0.5 \AA{} and a wavelength range of $\simeq$4050--7100 \AA{} were used. The polarimetric long-slit with a width of 1.5$^{''}$ was used. An exposure of an Argon lamp was taken after the science frames for wavelength calibration purposes.

\medskip
\noindent
The CCD pre-processing of the observations was carried out using the \textsc{polsalt-beta}\footnote{See  \url{https://github.com/saltastro/polsalt/} for more details.} software \citep{2003SPIE.4843..170N, 2012AIPC.1429..248N, 2016SPIE.9908E..2KP}, which is based on the \textsc{pysalt} package \citep{2010SPIE.7737E..25C}. The reduction steps included overscan correction, bias subtraction, and gain correction. Wavelength calibration for both the O and E beams was performed using Argon arc-lamp exposures obtained with the same instrumental configuration. The O and E beams of each spectrum were then extracted with \textsc{polsalt} following the procedures described in \cite{2020MNRAS.492.4298K}. The degree of circular polarisation (V/I) was computed using Equation (1) of \cite{2005A&A...442..651E} from two consecutive exposures, with the quarter-wave retarder plate rotated by $\pm$45$^{\circ}$. The results are shown in Fig. \ref{fig:Spectropolarimetric} and will be discussed in Section \ref{sec:spectropol_discussion}.

\section{Results and Analysis}
\subsection{Photometry}
\subsubsection{TESS}\label{sec:TESS_phot}

\begin{figure*}
    \centering
    \includegraphics[width=0.9\linewidth]{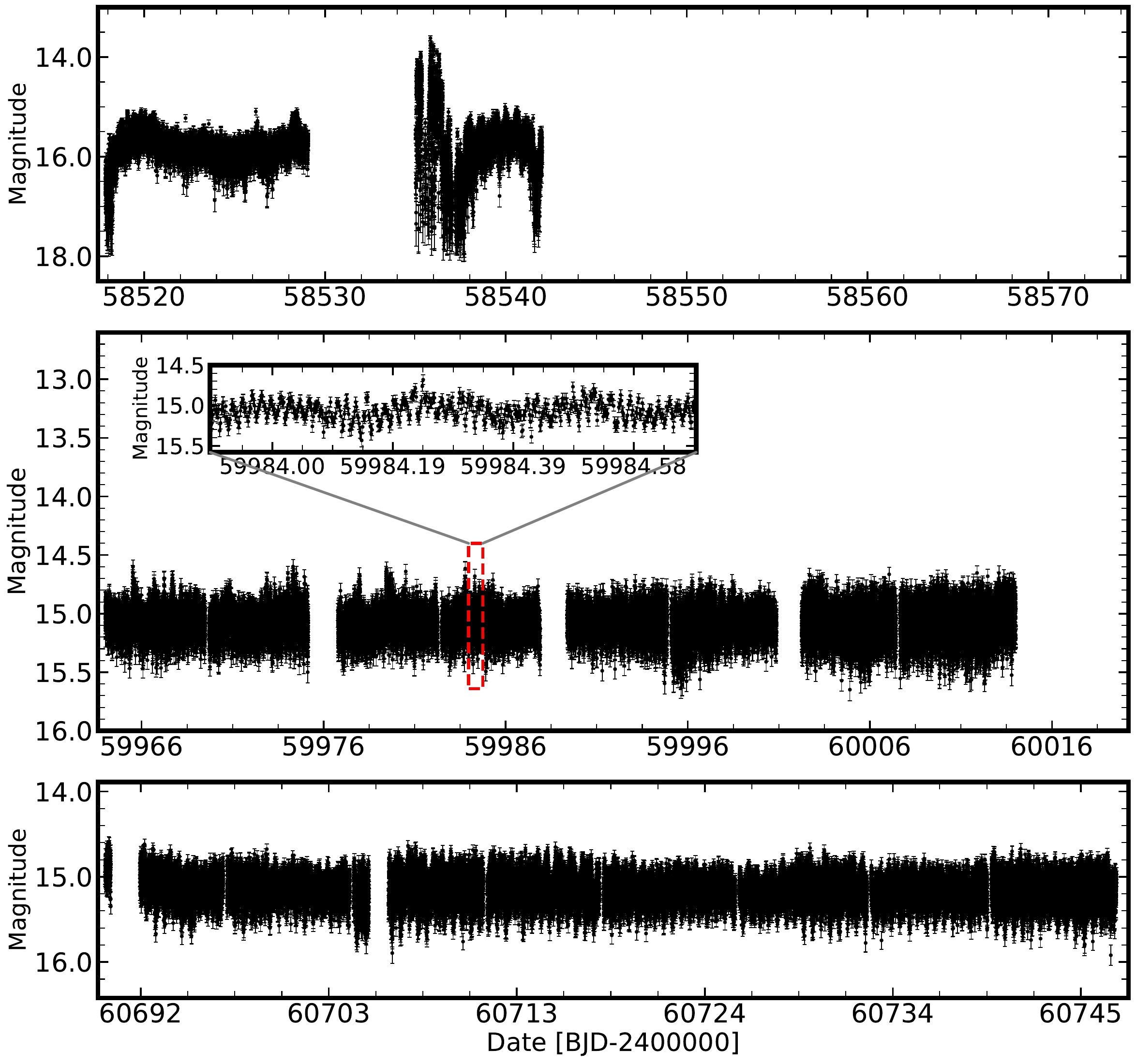}
    \caption{TESS light curve of 1RXS J080114.6–462324. The top panel shows data from Sector 8; the middle panel shows Sectors 61 and 62, including a zoomed-in region from [BJD – 2,400,000] = 59983.90 to 59984.68 that highlights a short and long periodic variability. The bottom panel presents the light curve from Sectors 88 and 89.}
    \label{fig:TESS_2019_2025}
\end{figure*}

\begin{figure*}
    \centering
    \includegraphics[width=1\linewidth]{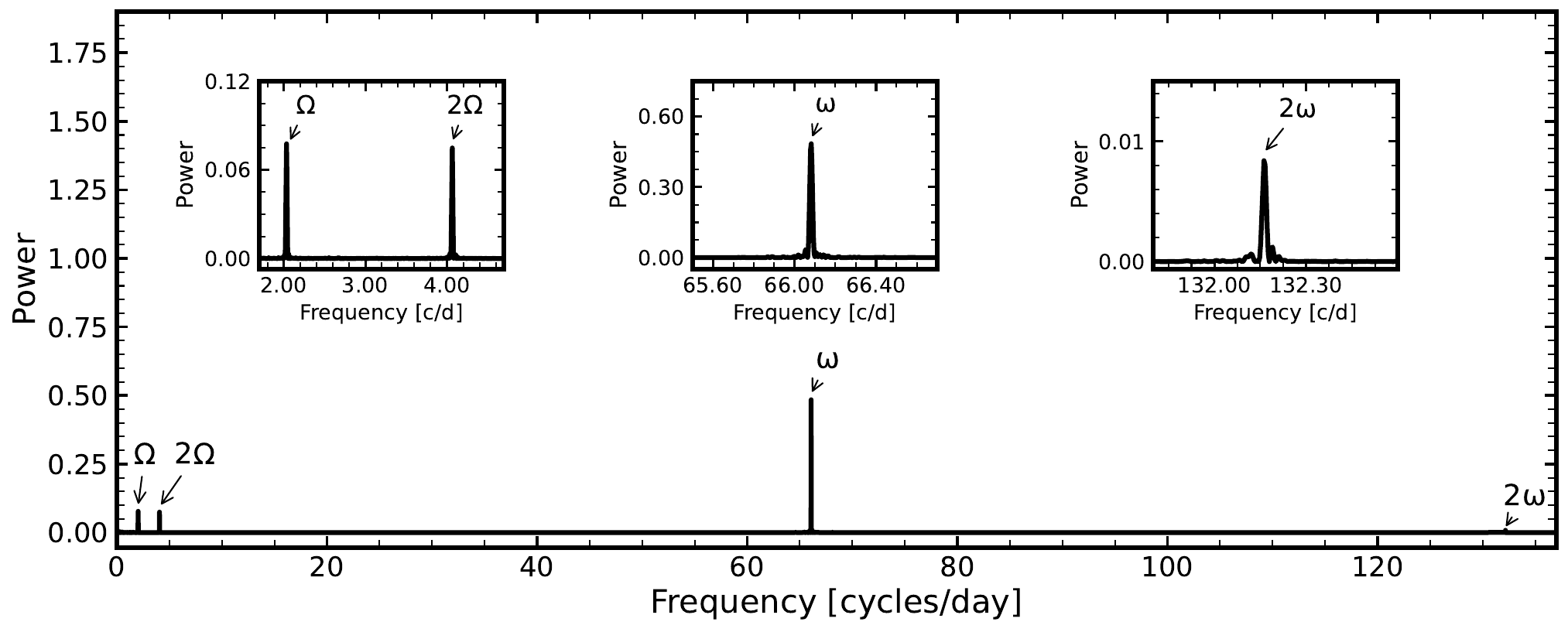}
    \caption{Lomb–Scargle periodogram of 1RXS J080114.6–462324, based on TESS photometric data from Sectors 88 and 89. The main panel shows the full power spectrum, with the symbols $\Omega$, $2\Omega$, $\omega$, and $2\omega$ indicating the orbital frequency, its second harmonic, the WD spin frequency, and its second harmonic, respectively. The insets highlight (left) the orbital frequency and its second harmonic, (middle) the spin frequency, and (right) the spin second harmonic. Revealing a prominent WD spin period of 1307.517 s and a possibly orbital period of approximately 11.803 h for the binary system.}
    \label{fig:LS_TESS_S88and89}
\end{figure*}

Fig. \ref{fig:TESS_2019_2025} presents the TESS light curves of 1RXS J080114.6–462324. The top panel shows a sudden increase in brightness, discussed in Section \ref{sec:photometry_discussion}. The zoomed-in region in the middle panel reveals clear variability, indicative of both short and relatively long periodic signals. 
The combined data from Sectors 88 and 89 were analysed using Lomb–Scargle periodogram \citep[see][]{1976Ap&SS..39..447L, 1982ApJ...263..835S}, to search for periodic signals, and the resulting periodogram is shown in Fig. \ref{fig:LS_TESS_S88and89}. Four prominent periodic signals were identified: two at lower frequencies, 2.03 cycles d$^{-1}$ and 4.06 cycles d$^{-1}$, and two at higher frequencies, 66.08 cycles d$^{-1}$ and 132.16 cycles d$^{-1}$. These signals have been reported previously. For example, \citet{2017MNRAS.470.4815B}, \citet{2018AJ....155..247H} and \citet{2024MNRAS.530.3974I} identified the 66.08 cycles d$^{-1}$ signal as the WD's spin frequency. In addition, \citet{2024MNRAS.530.3974I} reported a frequency at 4.06 cycles d$^{-1}$ as a possible binary orbital period and attributed the 132.16 cycles d$^{-1}$ signal to the second harmonic of the spin frequency. More recently, \citet{2025ApJS..279...48B} reported the 2.03 cycles d$^{-1}$ signal and interpreted it as the likely orbital period of the system.

\begin{figure}
    \centering
    \includegraphics[width=1.\linewidth]{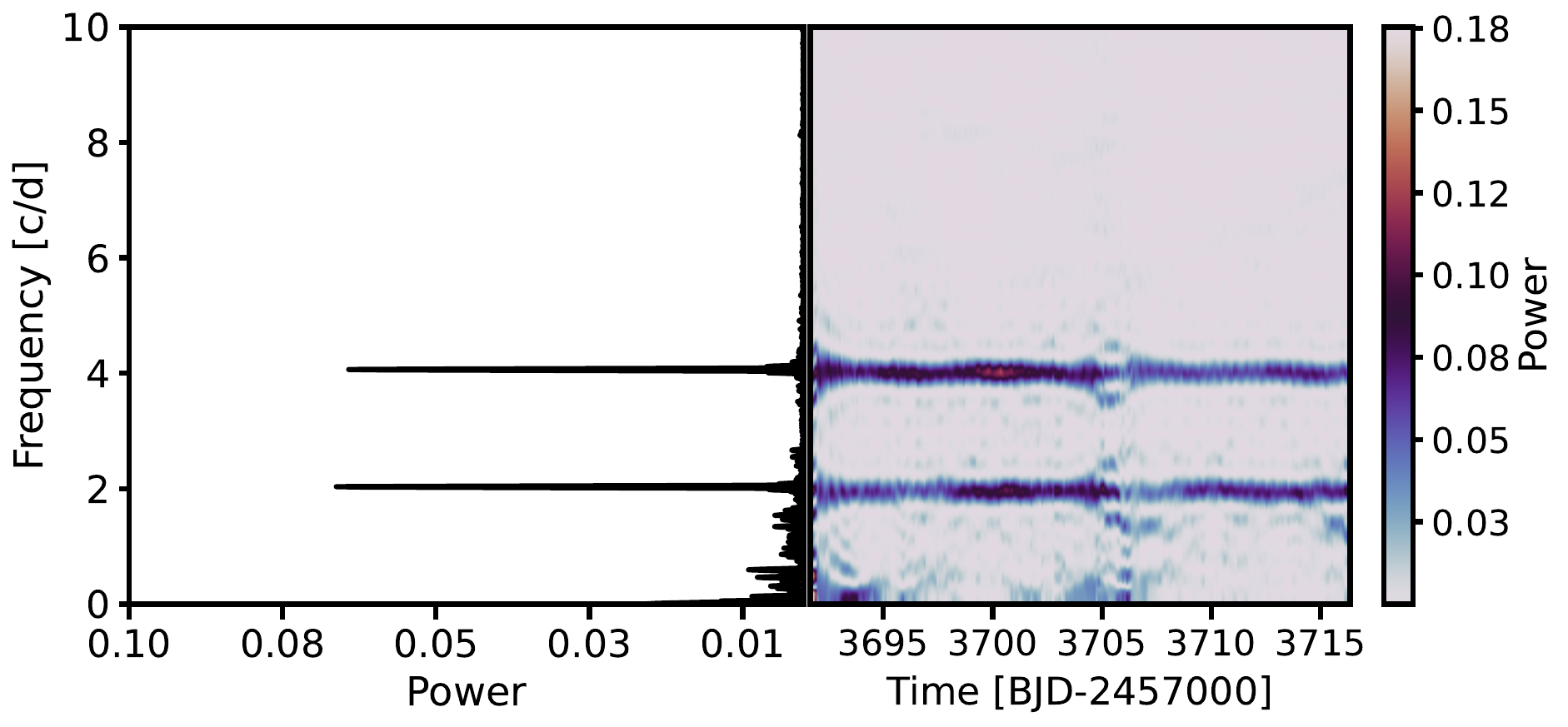}
    \caption{Dynamical Lomb–Scargle trailed spectra of TESS Sectors 88 and 89 for 1RXS J080114.6–462324, generated using a sliding 3-day window. Revealing a periodic signal at the frequency of 2.03 cycles d$^{-1}$ and 4.06 cycles d$^{-1}$. }
    \label{fig:Dynamical_Lombscargle}
\end{figure}

\begin{figure}
    \centering
    \includegraphics[width=\linewidth]{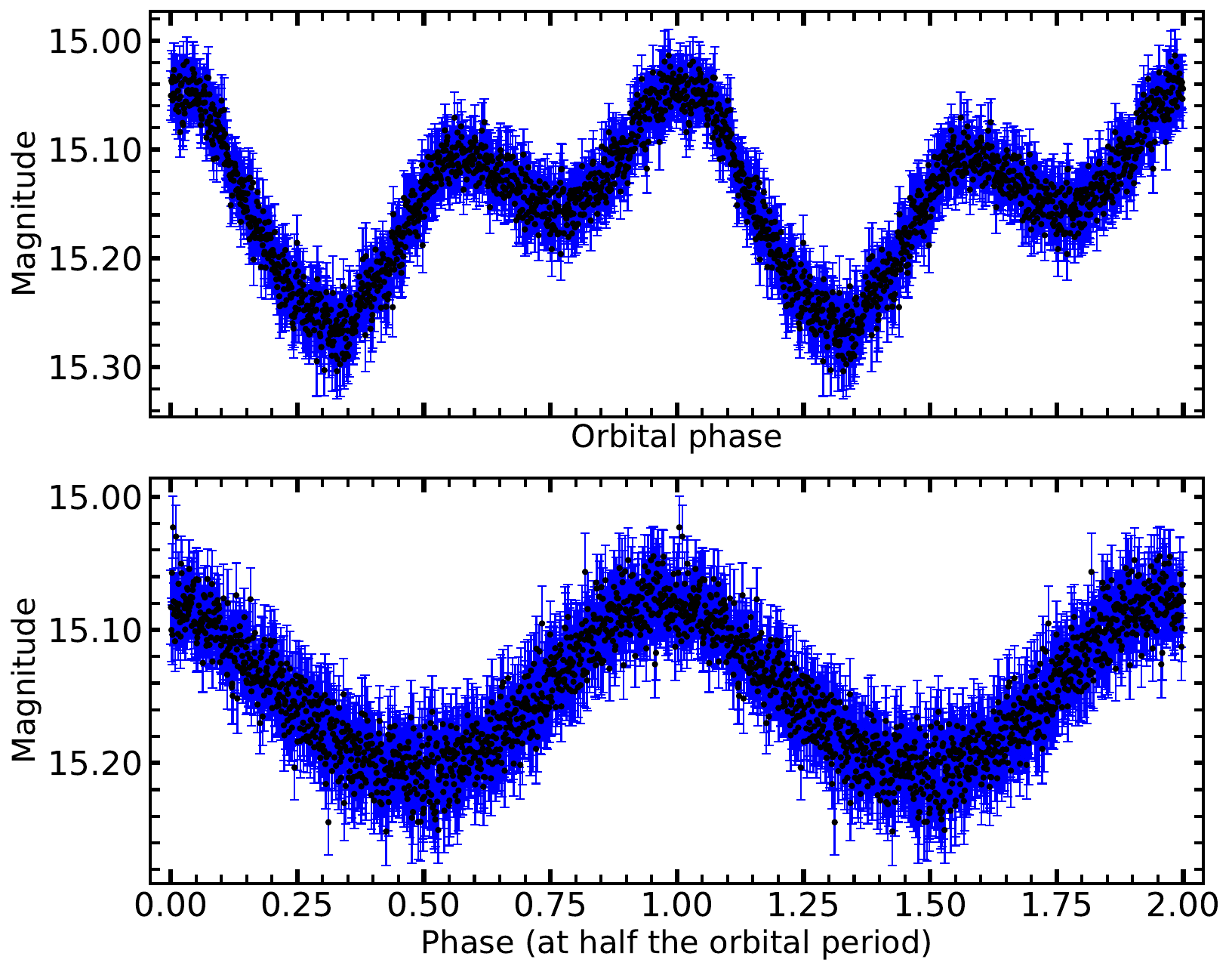}
    \caption{Phase-folded TESS light curves of 1RXS J080114.6–462324 from Sector 88 and 89 folded on frequencies of 2.0336 cycles d$^{-1}$ (top panel) and 4.0644 cycles d$^{-1}$ (bottom panel).}
    \label{fig:Combined_Folded}
\end{figure}

\medskip
\noindent
We applied a bootstrap resampling technique to the light curves to estimate the mean frequencies and associated uncertainties of these four detected signals. Additionally, a dynamical trailed spectrum was generated from the photometric light curve to investigate the temporal stability of the peaks detected at 2.03 cycles d$^{-1}$ and 4.06 cycles d$^{-1}$ in the Lomb–Scargle periodogram. This analysis was performed using a sliding 3-day window, within which the periodogram was computed for each segment. The individual power spectra were then stacked in chronological order to produce a two-dimensional time–frequency map, presented in Fig. \ref{fig:Dynamical_Lombscargle}. Finally, the combined TESS Sector 88 and 89 data were folded on the frequencies of 2.03 cycles d$^{-1}$ and 4.06 cycles d$^{-1}$, and the resulting phase-binned light curves are shown in the top and middle panels of Fig. \ref{fig:Combined_Folded}, showing two different light profiles. The choice of the orbital period, based on the different light profiles (see Fig. \ref{fig:Combined_Folded}), was further quantified using the combined standard deviation calculated from all bins of each of the folded light curves to evaluate which frequency produced the least scatter. The resulting standard deviations are $\sigma = 0.0244$ (for $\rm freq = 2.03 \; cycles \; d^{-1}$) and $\sigma = 0.0254$ (for $\rm freq = 4.06 \; cycles \; d^{-1}$), and these results are discussed in further details in Section \ref{sec:photometry_discussion}.

\subsubsection{SAAO 1.0-m and 1.9-m}

\begin{figure*}
    \centering
    \includegraphics[width=0.8\linewidth]{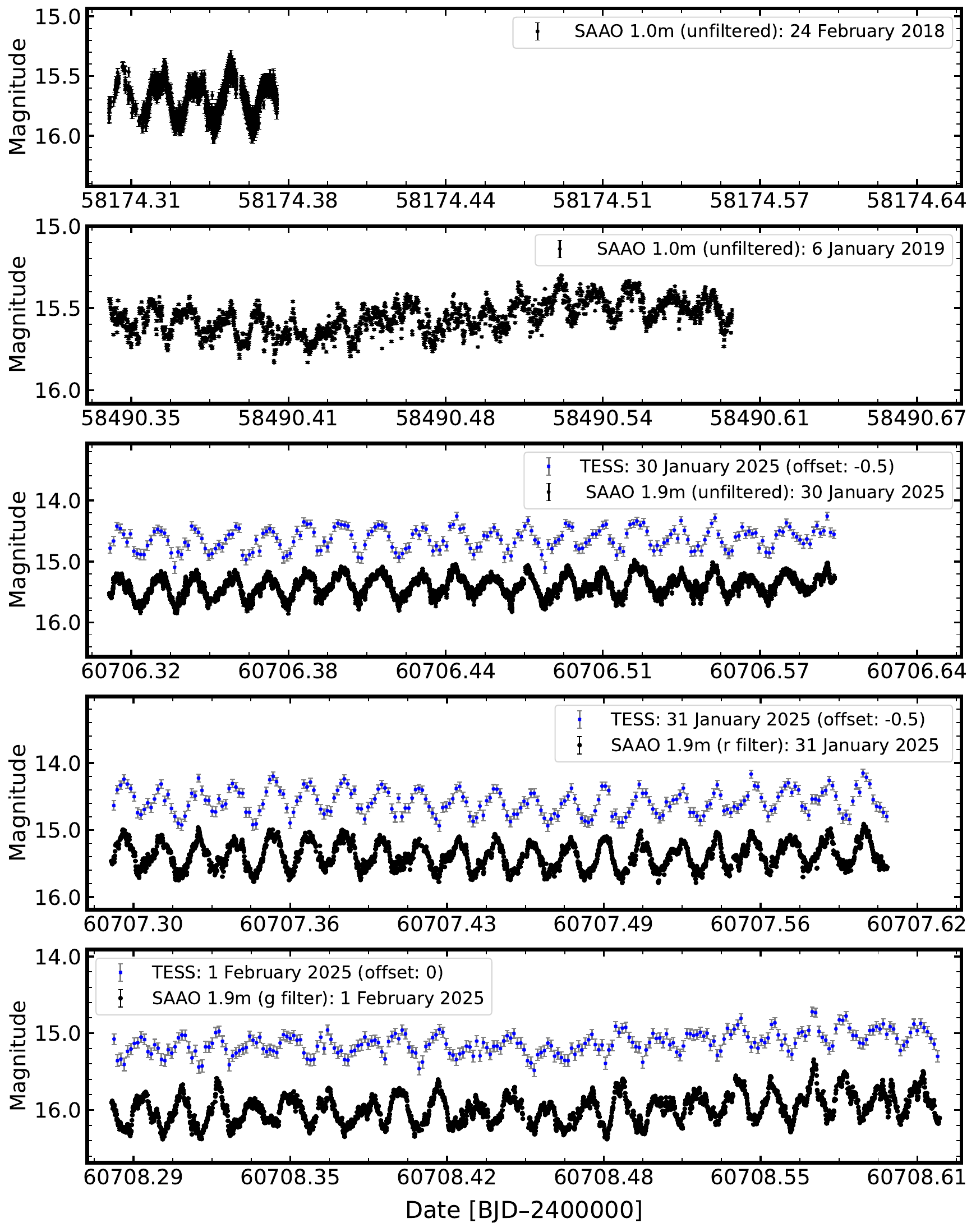}
    \caption{Photometric light curves of 1RXS J080114.6–462324. \textbf{Top panel}: Light curve obtained using the SAAO 1.0-m telescope. \textbf{Second panel}: Light curve obtained using the SAAO 1.9-m telescope. \textbf{Bottom three panels}: Simultaneous photometric light curves obtained over three nights using the SAAO 1.9-m telescope and TESS. Of these, the top panel shows unfiltered observations from the SAAO 1.9-m telescope alongside TESS data in the optical/near-infrared band. The middle panel shows r-band (red) photometry with simultaneous TESS observations. The bottom panel shows g-band (blue) photometry with simultaneous TESS observations.}
    \label{fig:1m_1.9m_tess}
\end{figure*}

\begin{figure*}
    \centering
    \includegraphics[width=0.9\linewidth]{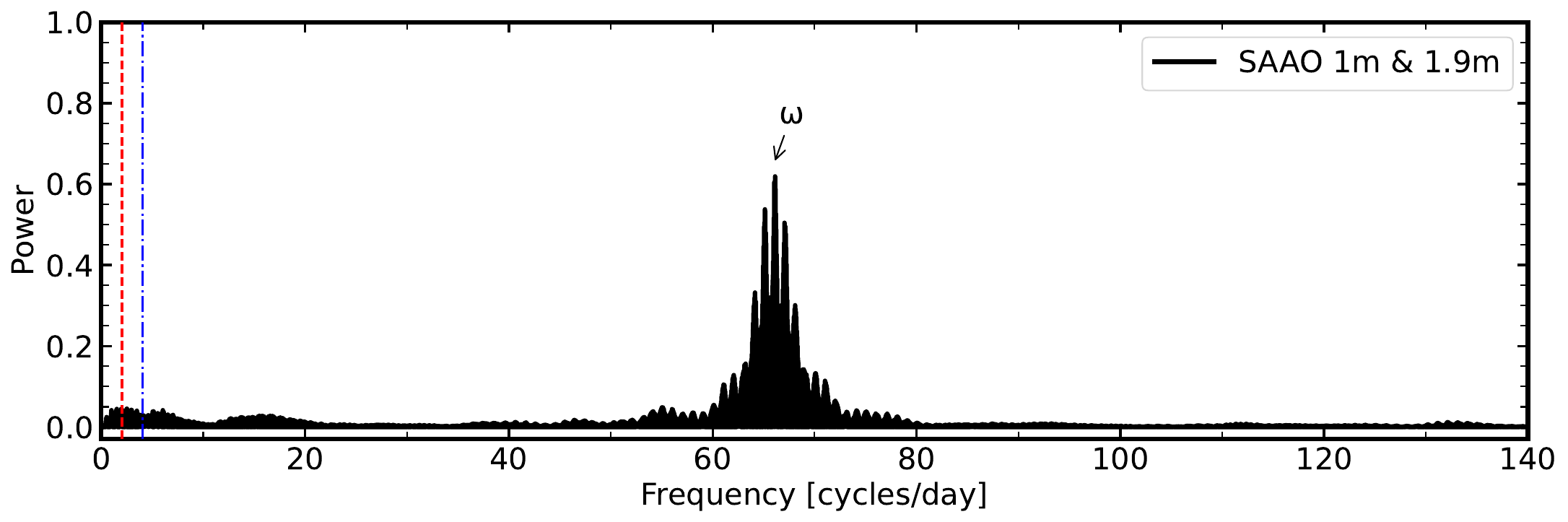}
    \caption{The Lomb-Scargle periodogram of the mean-subtracted light curve of 1RXS J080114.6–462324 derived from the combined data from the SAAO 1.0-m and 1.9-m telescopes.}
    \label{fig:SAAO_periodogram}
\end{figure*}

Fig. \ref{fig:1m_1.9m_tess} shows the light curves of 1RXSJ080114.6–462324 obtained with the SAAO 1.0-m and 1.9-m telescopes, together with simultaneous observations from TESS Sector 88. A clear short periodic signal is evident; we then examined the SAAO 1.0-m and 1.9-m mean-subtracted light curves using Lomb–Scargle periodogram. The resulting periodogram is shown in Fig. \ref{fig:SAAO_periodogram}. In contrast to the TESS data, these observations reveal a single prominent periodic signal at a frequency of 66.08 cycles d$^{-1}$ (indicated by $\omega$). For comparison, the dashed and dash–dot lines mark the locations of the 2.03 cycles d$^{-1}$ and 4.06 cycles d$^{-1}$ peaks identified in the TESS data, which are absent in our dataset. This non-detection is likely due to the shorter observational baseline relative to TESS, as well as data gaps and sampling effects that introduce strong window function artifacts \citep[e.g.,][]{2015A&A...574A..18P,2015A&A...575A..78P}. This is further discussed in Section \ref{sec:photometry_discussion}.

\noindent

\subsubsection{The Spin and Orbital Ephermeris}
\begin{figure*}
    \centering
    \includegraphics[width=0.8\linewidth]{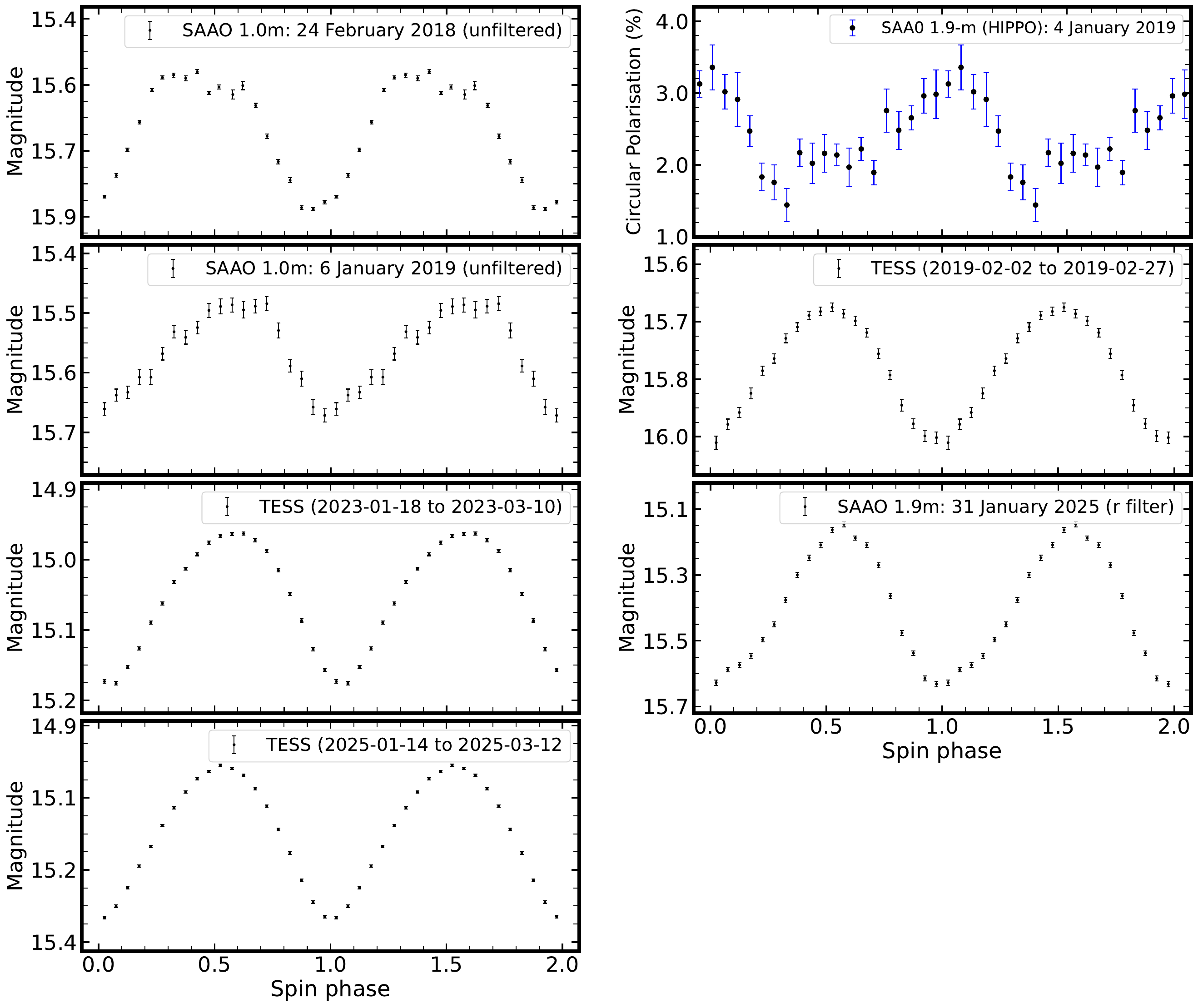}
    \caption{Spin-phase-binned photometric and polarimetric observations of 1RXS J080114.6$-$462324. \textbf{Top row:} Optical light curves obtained with the SAAO 1.0-m telescope using the SHOC instrument (left), and circular polarimetry obtained with the HIPPO instrument (right). \textbf{Second row:} Optical light curves from the SAAO 1.0-m telescope using SHOC (left), and TESS light curves from Sector 8 (right). \textbf{Third row:} Combined TESS light curves from Sectors 61 and 62 (left), and optical light curves from the SAAO 1.9-m telescope using the SHOCnWOnder instrument (right). \textbf{Bottom panel:} Combined TESS light curves from Sectors 88 and 89.}
    \label{fig: Phased_LC}
\end{figure*}

The mean-subtracted TESS light curves from Sectors 8, 61, 62, 88, and 89 were combined and analysed using the Lomb-Scargle periodogram, from which the best-fitting frequencies were determined. Based on our analysis, the signal at 2.03 cycles d$^{-1}$ was identified as the binary orbital frequency, with the 4.06 cycles d$^{-1}$ signal corresponding to its second harmonic. The frequency at 66.08 cycles d$^{-1}$ was attributed to the WD's spin period, with its harmonic detected at 132.16 cycles d$^{-1}$. These two periods were then used to derive the spin and orbital ephemerides of 1RXS J080114.6--462324, as presented in Eqs. \ref{spin_ephemeris} and \ref{orbital_ephemeris}. We also report refined values of $1307.5179 \pm 0.0527$ s for the spin period of the WD and $11.8026 \pm 0.0004$ h for the orbital period of the system.

\begin{equation}\label{spin_ephemeris}
    T_{\text{spin}, n} (\mathrm{BJD}) = 2458486.91332(6) + 0.01513331(61)\,n
\end{equation}
\begin{equation}\label{orbital_ephemeris}
    T_{\text{orb}, n} (\mathrm{BJD}) = 2458486.86414(6) + 0.491777(15)\,n
\end{equation}
\noindent
The epoch of the spin ephemeris was determined by folding the light curve at the WD’s spin period and identifying the time (in BJD) corresponding to the minimum flux. Similarly, the epoch of the orbital ephemeris was determined by folding the light curve at the orbital period and locating the time (in BJD) of maximum brightness associated with orbital modulation. The values in parentheses indicate the uncertainties in the final digits. The light curves from the SAAO 1.0-m, 1.9-m, and TESS were subsequently folded using the spin ephemeris given in Eq. \ref{spin_ephemeris}, and the resulting spin-phase-binned light curve is presented in Fig. \ref{fig: Phased_LC}. The error bars indicate the standard deviation of the binned flux values.

\subsection{Spectroscopy}\label{sec:spec_result}
\begin{figure}
    \centering
    \includegraphics[width=1\linewidth]{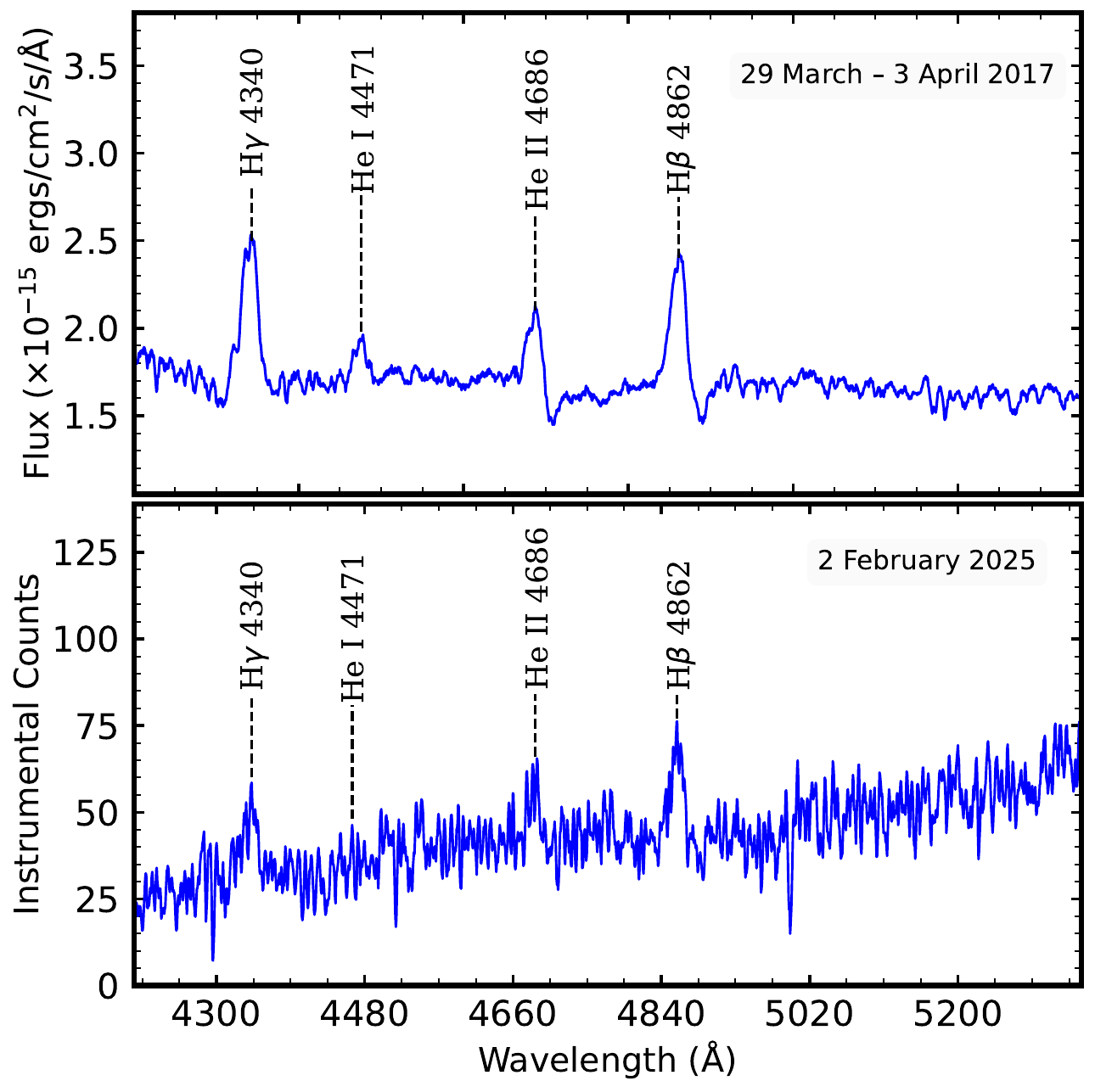}
    \caption{Optical spectra covering the wavelength range 4200--5400\,\AA, obtained with the SAAO 1.9-m telescope. \textbf{Top panel}: shows the flux-calibrated average spectrum combined over six observing nights (29 March to 3 April 2017). \textbf{Bottom panel}: presents the average (non-flux-calibrated) spectrum from 2 February 2025. Prominent emission lines are identified, including H$\gamma$, He\,\textsc{i} (4471\,\AA), He\,\textsc{ii} (4686\,\AA), and H$\beta$.}
    \label{fig:average_spectrum}
\end{figure}

\begin{figure}
    \centering
    \includegraphics[width=1\linewidth]{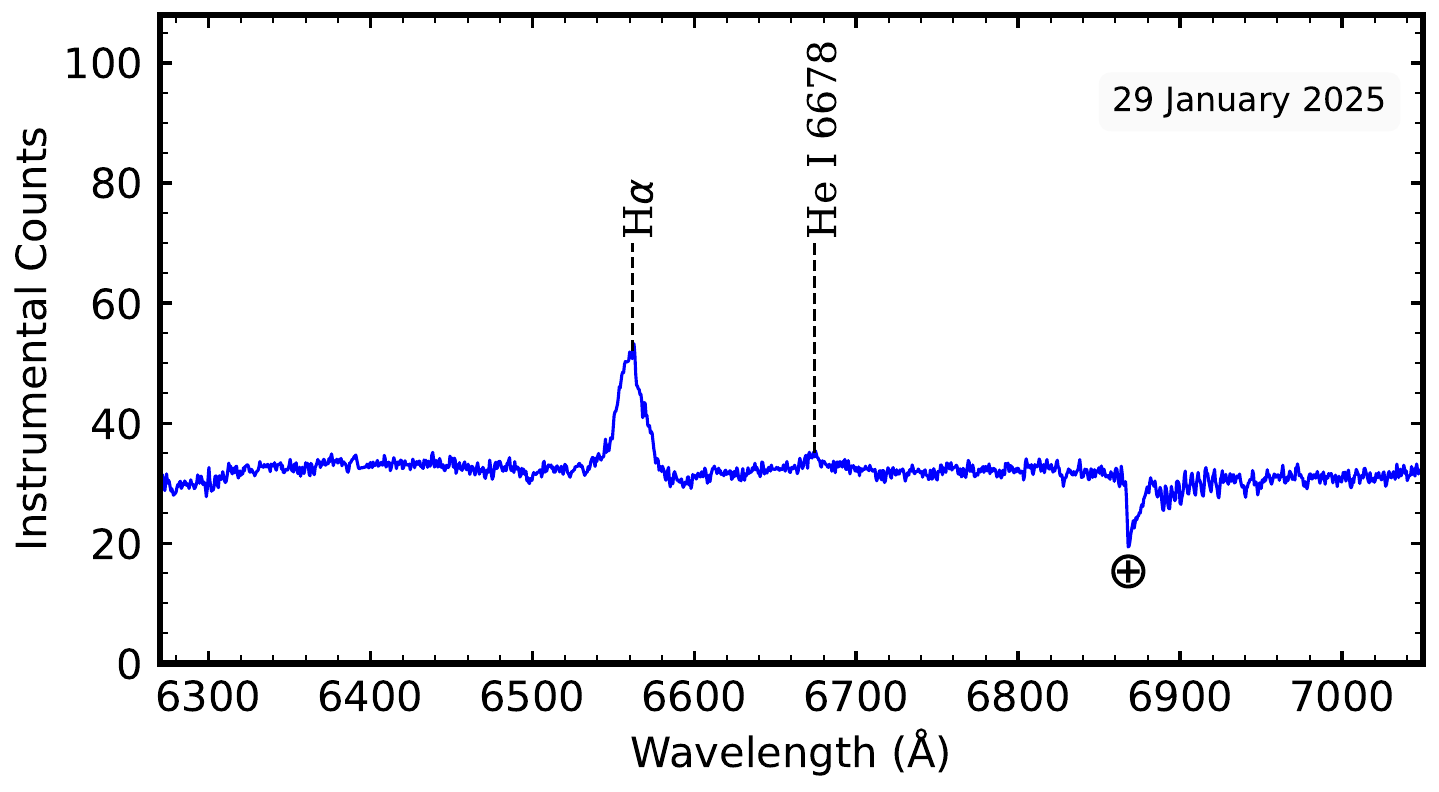}
    \caption{Optical spectra covering the wavelength range 6270--7050\,\AA, obtained with the SAAO 1.9-m telescope on 2025 January 29. Prominent emission lines are identified, including H$\alpha$ and He\,\textsc{i} (6678\,\AA). The $\oplus$ symbol marks regions affected by telluric absorption.}
    \label{fig:H_alpha}
\end{figure}

Fig. \ref{fig:average_spectrum} shows the time-averaged spectra of 1RXSJ080114.6–462324 obtained with the SAAO 1.9-m telescope using grating 4, while Fig. \ref{fig:H_alpha} presents the spectrum acquired with grating 5. 
The blue spectra exhibit the characteristic emission lines of IPs, including the hydrogen Balmer lines H$\beta$ and H$\gamma$, as well as He\,\textsc{i} $\lambda4471$ and  He\,\textsc{ii} $\lambda4686$, the latter is associated with energetic accretion processes. 
The red spectrum shown in Fig. \ref{fig:H_alpha} also displays the H$\alpha$ emission line and He\,\textsc{i} $\lambda6678$, a further indicator of ongoing accretion. These features are consistent with the spectrum of 1RXS J080114.6–462324 first presented by \citet{2010A&A...519A..96M}, which showed Balmer emission up to H$\beta$ alongside helium emission lines. 
However, we also identify absorption dips in the time-average spectra of 1RXS J080114.6–462324 (see Fig. \ref{fig:average_spectrum}) that were not reported by \citet{2010A&A...519A..96M}, likely due to their lower spectral resolution or different state of the system. These features are discussed further in Section \ref{sec:spectroscopy_discussion}.

\begin{figure*}
    \centering
    \includegraphics[width=\linewidth]{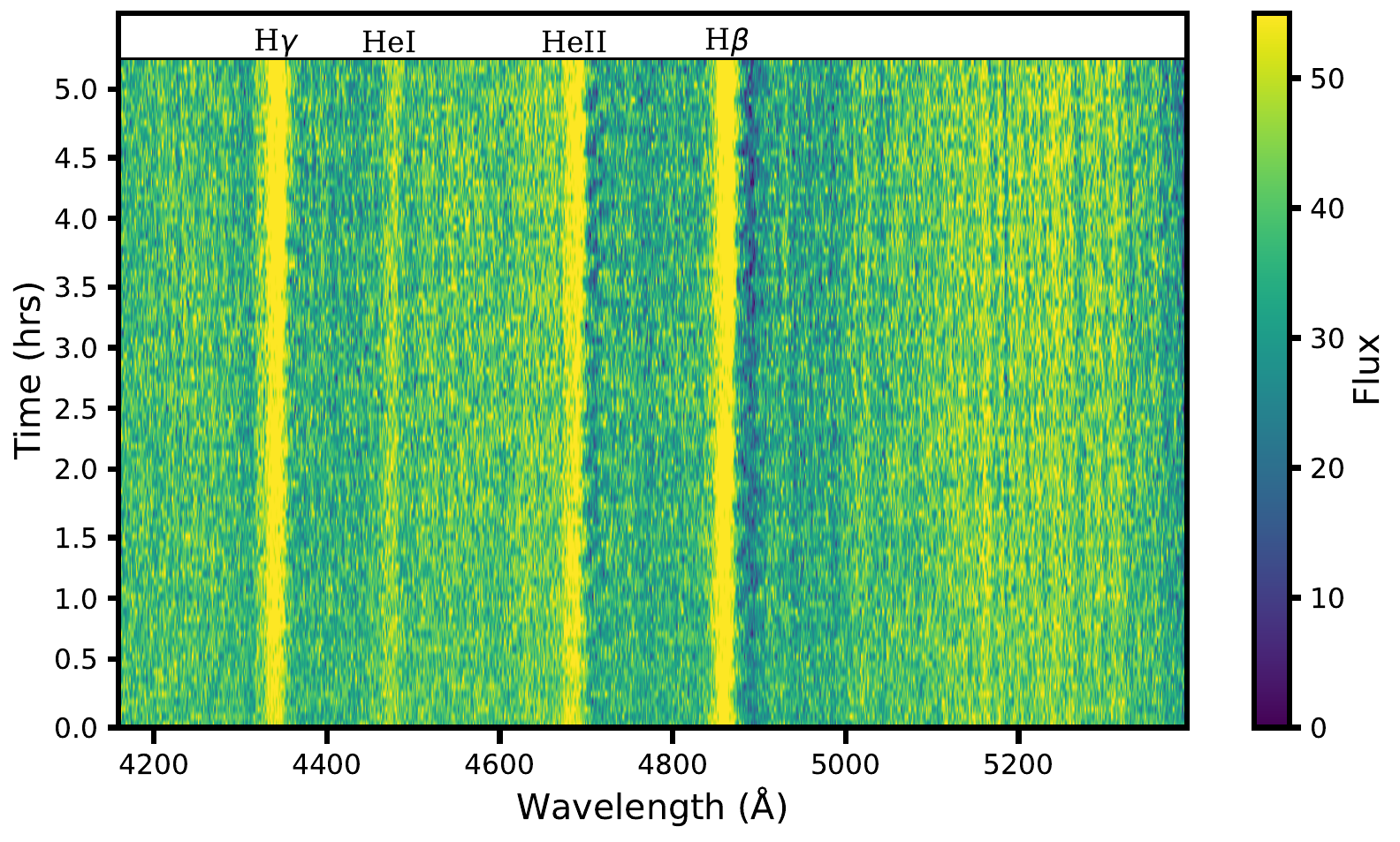}
    \caption{Continuum-subtracted trailed spectra obtained from spectroscopic observations on 3 April 2017. Prominent emission lines are labeled. A distinct redshifted absorption component is evident in the He \textsc{ii} and H$\beta$ lines.}
    \label{fig:trailed_spectra}
\end{figure*}

\begin{figure}
    \centering
    \begin{subfigure}{0.5\textwidth}
        \centering
        \includegraphics[width=\linewidth]{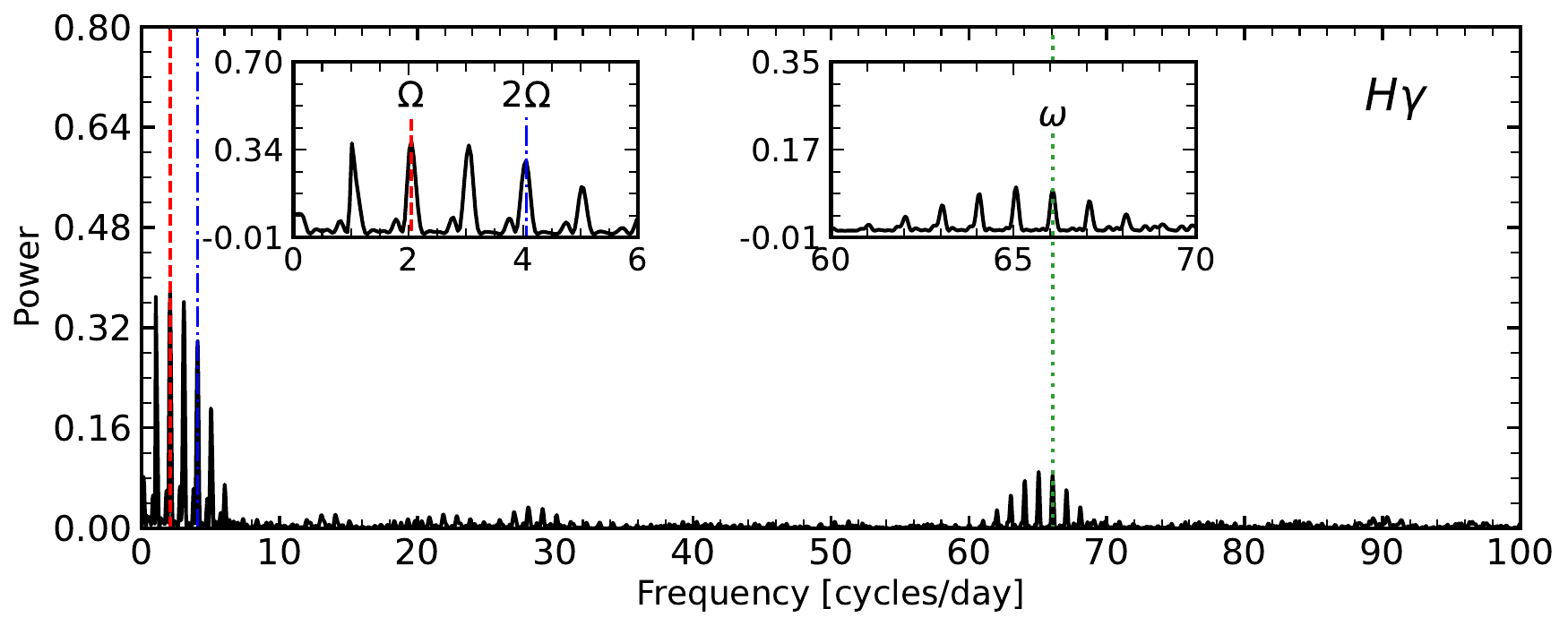}
    \end{subfigure}
    \begin{subfigure}{0.5\textwidth}
        \centering
        \includegraphics[width=\linewidth]{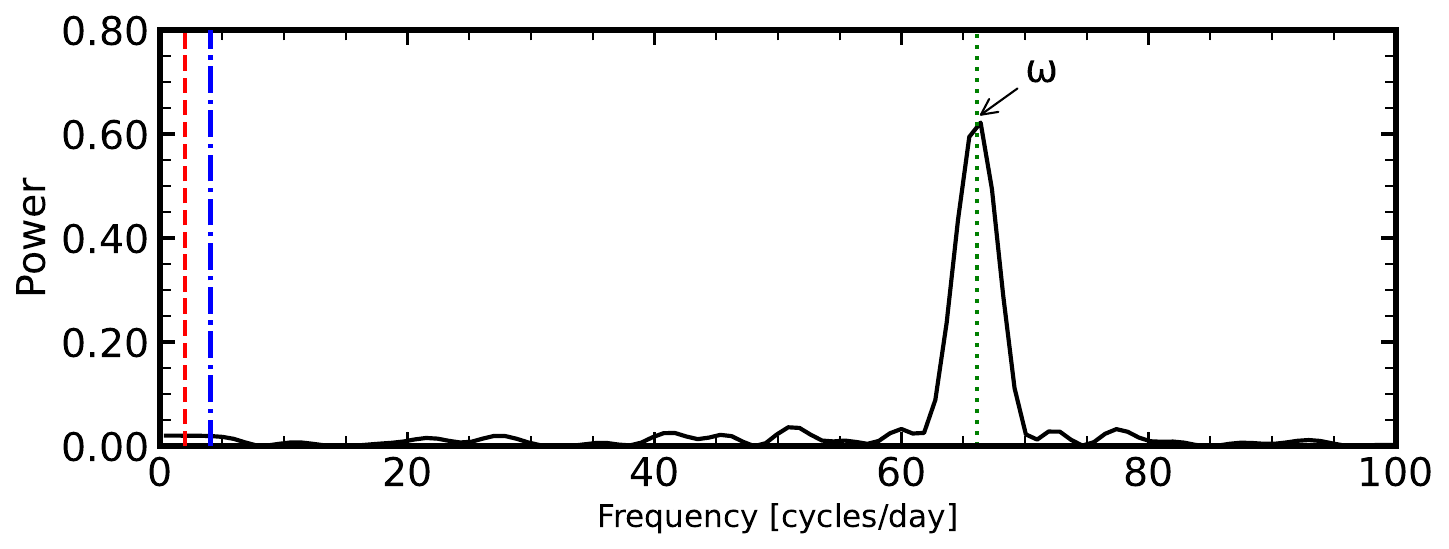}
    \end{subfigure}
    \caption{The L-S periodogram of the spectroscopic data. \textbf{Top panel:} Lomb-Scargle periodogram of the central value from a Gaussian fit to the H$\gamma$ emission line, derived from combined spectroscopic data obtained over six nights of observations. The blue dash-dotted line marks the proposed orbital frequency (2.03 cycles d$^{-1}$), also indicated in the inset on the left. The green dotted line denotes the WD spin frequency ($\omega$), measured at 66.081 cycles d$^{-1}$, and highlighted in the zoomed-in inset on the right. \textbf{Bottom panel: } The L-S periodogram of the central value from a Gaussian fit to the redshifted absorption dip next to the H$\beta$ emission line}
    \label{fig:Gamma_LS}
\end{figure}

\medskip
\noindent
From the spectra taken on 3 April 2017 with grating 4, we constructed a trailed spectrum covering the full wavelength range; the results are shown in Fig. \ref{fig:trailed_spectra}. We identified short-term radial velocity variations as well as a prominent red-shifted absorption feature adjacent to the He\,\textsc{ii} $\lambda4686$, and H$\beta$. 
To search for any periodic modulations in the trailed spectra, we fitted a Gaussian profile to the H$\gamma$ emission line in each time-resolved spectrum for all six nights of observations (2017 March 29–April 3). The fitted Gaussian central wavelength provided a time series of the emission line centroid. We then applied a Lomb–Scargle periodogram analysis to this time series to examine how the centroid varied over time. The resulting periodogram is presented in the top panel of Fig. \ref{fig:Gamma_LS}, illustrating the dominant periodicities in the data. A strong periodic signal is detected at $\sim$66.08 cycles d$^{-1}$, corresponding to the spin period of the WD. Two additional signals, at 2.03 cycles d$^{-1}$ and 4.06 cycles d$^{-1}$, are also present; in line with the frequencies detected in the TESS light-curve. This technique was further applied to the observed red-shifted dip adjacent to H$\beta$, in order to extract the time series of the absorption dip centroid. For the absorption component, we selected the spectra obtained on 3 April 2017, as we could easily fit the absorption component without any loss of information. Although the feature is also present on other nights during 2017, it is slightly less pronounced. The resulting periodogram (Fig. \ref{fig:Gamma_LS}) also shows a strong periodic signal at $\sim$66.08 cycles d$^{-1}$.

\subsection{Polarimetry}\label{sec:results_polarimetry}

\begin{figure*}
    \centering
    \includegraphics[width=0.8\linewidth]{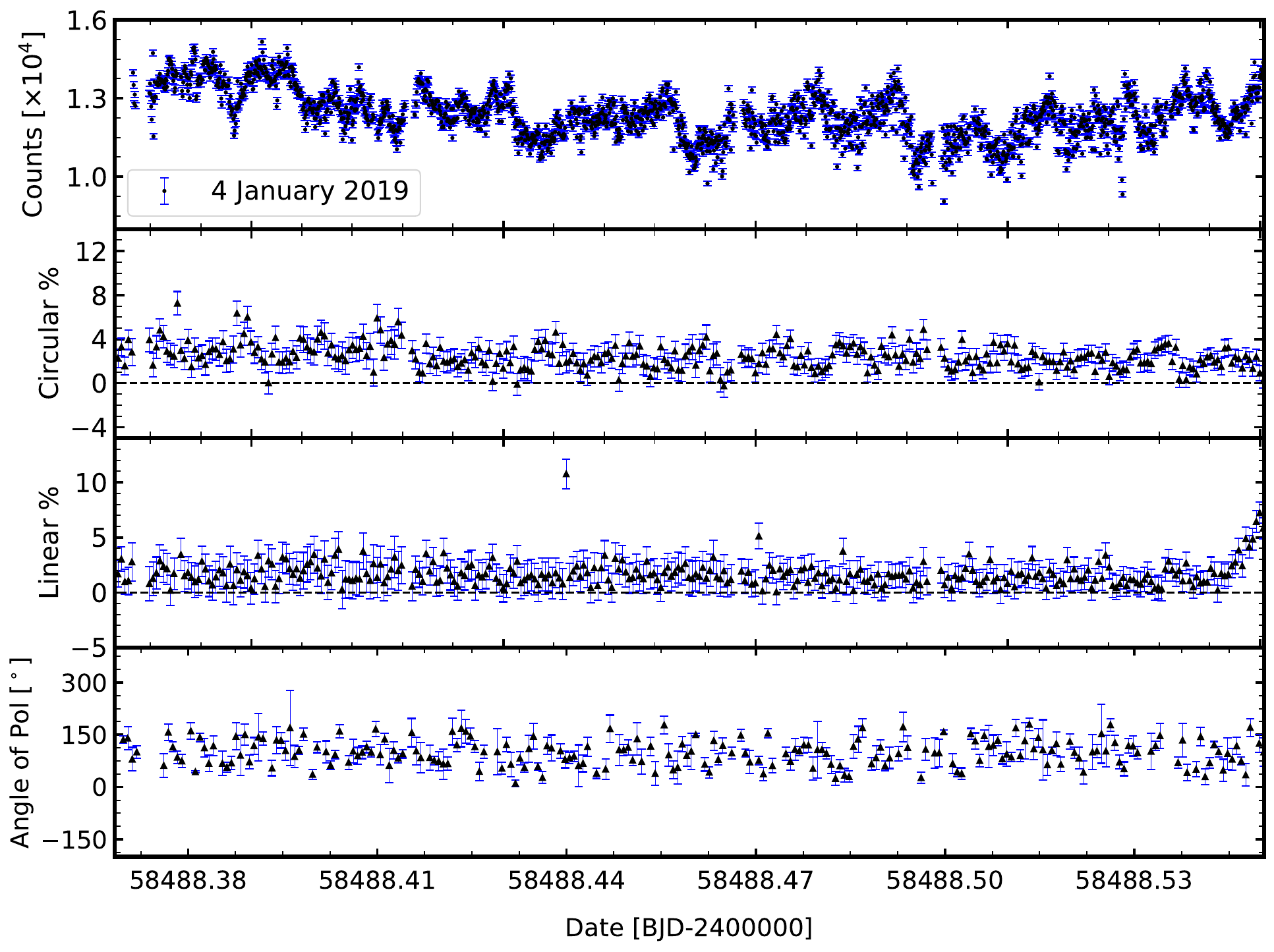}
    \caption{Photometric and polarimetric observations obtained on 2019 January 4 using a clear filter. The top panel shows the photometric light curve, binned in 10-second intervals. The second, third, and fourth panels present the circular and linear polarisation and position angle, respectively, binned in 60-second intervals.}
    \label{fig:Photopolarimetric}
\end{figure*}

Fig. \ref{fig:Photopolarimetric} shows the HIPPO photometric and polarimetric light curve obtained on 2019 January 4 using a clear filter. The top panel shows the photometric light curve with brightness given as total counts, showing some variability over the observing period. The second panel presents the percentage of circular polarisation which reveals a clear signature of circularly polarised emission between $\sim$0-5\% with short variability timescales similar to the photometry. The third panel presents the percentage of linearly polarisation emission, which appears at a constant level of generally less than  $\sim$3\% with no apparent variability. The increase in linear polarisation towards the end of the dataset is contamination due to the beginning of twilight.  The bottom panel of Fig. \ref{fig:Photopolarimetric} shows the change in polarisation position angle over our observing period, also appearing flat, with the error bars indicating 1$\sigma$ uncertainties propagated from the measured Stokes $Q$ and $U$ parameters.

\begin{figure}
    \centering
    \includegraphics[width=1.02\linewidth]{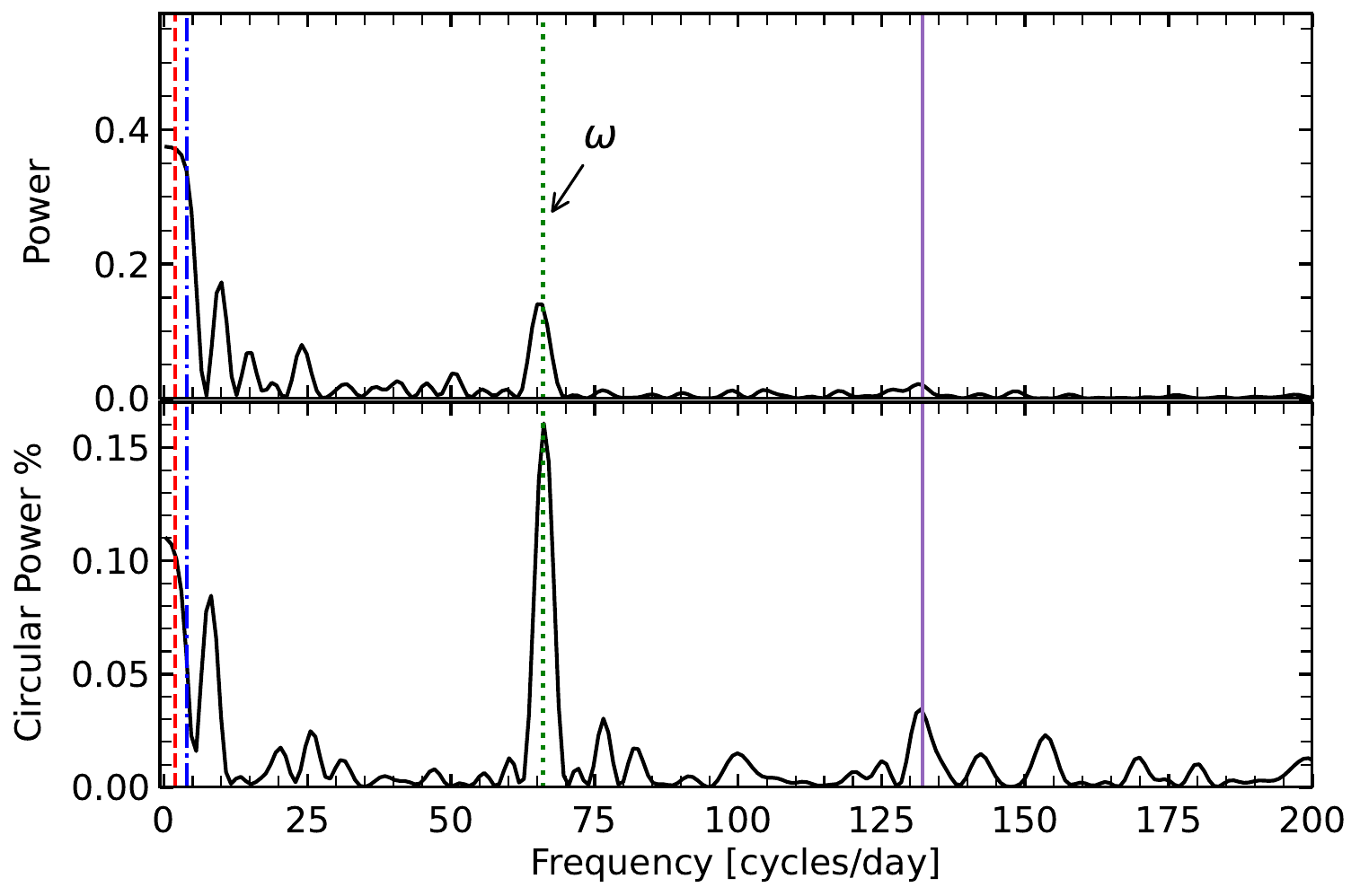}
    \caption{Lomb-Scargle periodogram of the photopolarimetry data observed on 2019 January 4. The power spectrum of the photometry data is shown in the top panel, and the circular polarised counts percentage power spectrum is shown in the bottom panel. In both power spectra, the red dashed and blue dash–dotted vertical lines mark the expected locations of the binary orbital frequency ($\Omega$) and its second harmonic ($2\Omega$), respectively. The green dotted and solid purple lines mark the locations of the spin frequency ($\omega$) and its second harmonic ($2\omega$), respectively.}
    \label{fig:PHOTO_POL_SCARGLE}
\end{figure}

\medskip
\noindent
We investigated the variability present in the photometric light curve and the circular polarisation emission shown in Fig. \ref{fig:Photopolarimetric} by performing Lomb-Scargle periodogram analyses on both datasets. The resulting periodograms are presented in Fig. \ref{fig:PHOTO_POL_SCARGLE}, with the photometric power spectrum shown in the top panel and the circular polarimetric percentage power spectrum in the bottom panel. In both the photometric and circular polarimetric data, the periodograms reveal prominent peaks at a frequency of $\sim$66.08 cycles d$^{-1}$. The dashed and dash–dot vertical lines indicate the expected positions of the periodic signals at 2.03 cycles d$^{-1}$ and 4.06 cycles d$^{-1}$, respectively, as observed in the TESS light curve periodogram. The $\sim66.08$ cycles d$^{-1}$ signal corresponds to the known spin period of the WD. This is a clear indication of a spin-modulated circular polarisation emission. Consequently, we folded the circular polarisation percentage light curve on the WD's spin period, and the resulting spin-phase-binned circular polarisation percentage light curve is shown in Fig. \ref{fig: Phased_LC}. These results are further discussed in Section \ref{sec:photopol_discussion}.

\subsection{Spectropolarimetry with SALT}

\begin{figure}
    \centering
    \includegraphics[width=1\linewidth]{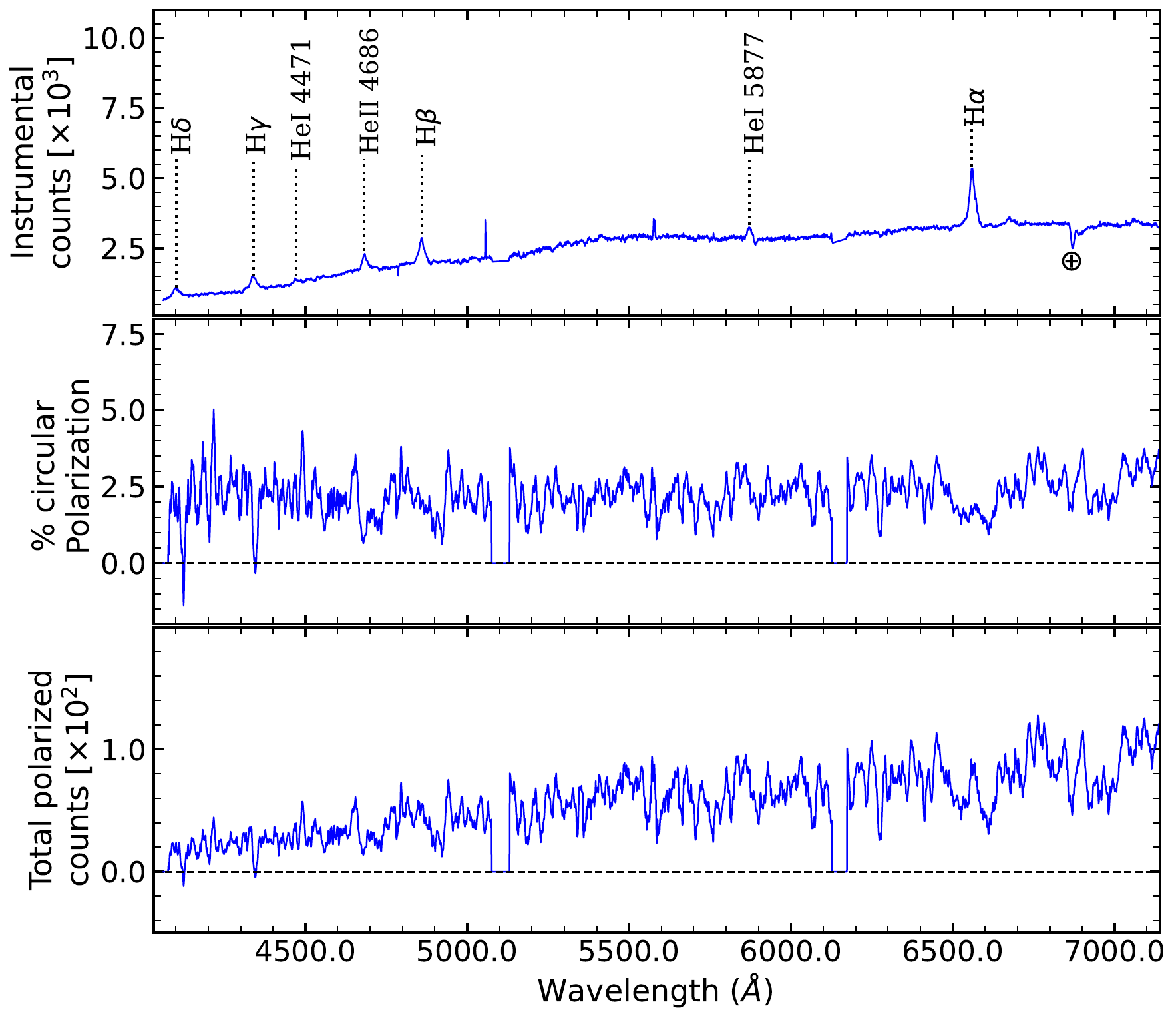}
    \caption{The spectra of 1RXSJ080114.6–462324 centred on the spin phase ($\phi  = 0.62$). \textbf{Top panel}: average spectrum with prominent emission lines identified. The $\oplus$ symbol marks regions affected by telluric (atmospheric) absorption. \textbf{Middle panel}: Circular polarisation percentage spectra. \textbf{Bottom panel}: Total polarised counts spectrum. The other two dips around 5100 \AA{} and 6150 \AA{} observed are chip gaps.}
    \label{fig:Spectropolarimetric}
\end{figure}

Fig. \ref{fig:Spectropolarimetric} shows the circular spectropolarimetric results of 1RXSJ080114.6–462324 obtained with SALT, centred on the spin-phase $\phi = 0.62$. The data show a clear signature of positive circularly polarised emission, with a maximum of about $\sim+5\%$\ within the wavelength range 4062–7155\,\AA. The level of circular polarisation does not seem to vary with the wavelength range but does vary with the spin phase. These results further support the detection of a circularly polarised emission in our photopolarimetry observations. Our spectropolarimetric observations show no evidence of cyclotron humps or Zeeman-splitting features, and this is further discussed in Section \ref{sec:spectropol_discussion}.

\section{Discussion}

We presented multi-instrument photometric, spectroscopic, and polarimetric observations of the IP 1RXSJ080114.6–462324. 

\subsection{Photometry}\label{sec:photometry_discussion}

Our analysis revealed a dominant periodic signal at approximately 66.08 cycles d$^{-1}$. 
The extended time coverage provided by the TESS observations further revealed frequencies at 2.03 cycles d$^{-1}$ and 4.06 cycles d$^{-1}$ in addition to frequencies at 66.08 cycles d$^{-1}$ and 132.16 cycles d$^{-1}$. 
The dynamical trailed spectra presented in Fig. \ref{fig:Dynamical_Lombscargle} show that the two low-frequency signals remain stable over time, exhibiting comparable power. The vertical artifacts in the dynamical trailed spectra around 3705–3707 [BJD-2457000] are attributable to gaps in the data. 
No orbital sidebands of the spin frequency were detected. We attribute the signal at 66.08 cycles d$^{-1}$, corresponding to $1307.4860 \pm 0.0527$s, to the spin period of the WD. These results are consistent with previous determinations \citep{2017MNRAS.470.4815B, 2024MNRAS.530.3974I}. The signal at 132.16 cycles d$^{-1}$ is identified as the second harmonic of the spin period ($2\omega$).

\medskip
\noindent
We confirm the periodic signal at 4.06 cycles d$^{-1}$, previously reported by \citet{2024MNRAS.530.3974I}, but interpret it here as the second harmonic of the binary orbital frequency. The strong peak at 2.03 cycles d$^{-1}$ is therefore identified as the likely orbital frequency of the system, corresponding to a period of $11.8026 \pm 0.0004$\,h. 
This same period was recently reported by \citet{2025ApJS..279...48B}. Folding the TESS data on this period reveals a typical asymmetric double humped morphology for the light curve, indicating that 11.80\,h corresponds to the binary orbital period, while the period of 5.90\,h represents its second harmonic. 
In contrast, folding the data on the 5.90\,h harmonic (4.06 cycles d$^{-1}$) produces a broader profile with greater scatter, consistent with this signal being the second harmonic of the orbital period. The deviation of the data points from the mean can also be assessed through the calculated standard deviations. The slightly larger deviation observed for 4.06 cycles d$^{-1}$ ($\sigma = 0.0254$) compared to 2.03 cycles d$^{-1}$ ($\sigma = 0.0244$) indicates that the latter exhibits a more coherent phase-folded light curve, further supporting 2.03 cycles d$^{-1}$ ($P_{\rm orb} = 11.80$ hr) as the most probable binary orbital frequency. A similar light profile was reported by \citet{2025AJ....169..269J} for the IP IGR J14091–6108, where the observed double-peaked modulation was attributed to ellipsoidal variations.

\begin{figure}
    \centering
    \includegraphics[width=1\linewidth]{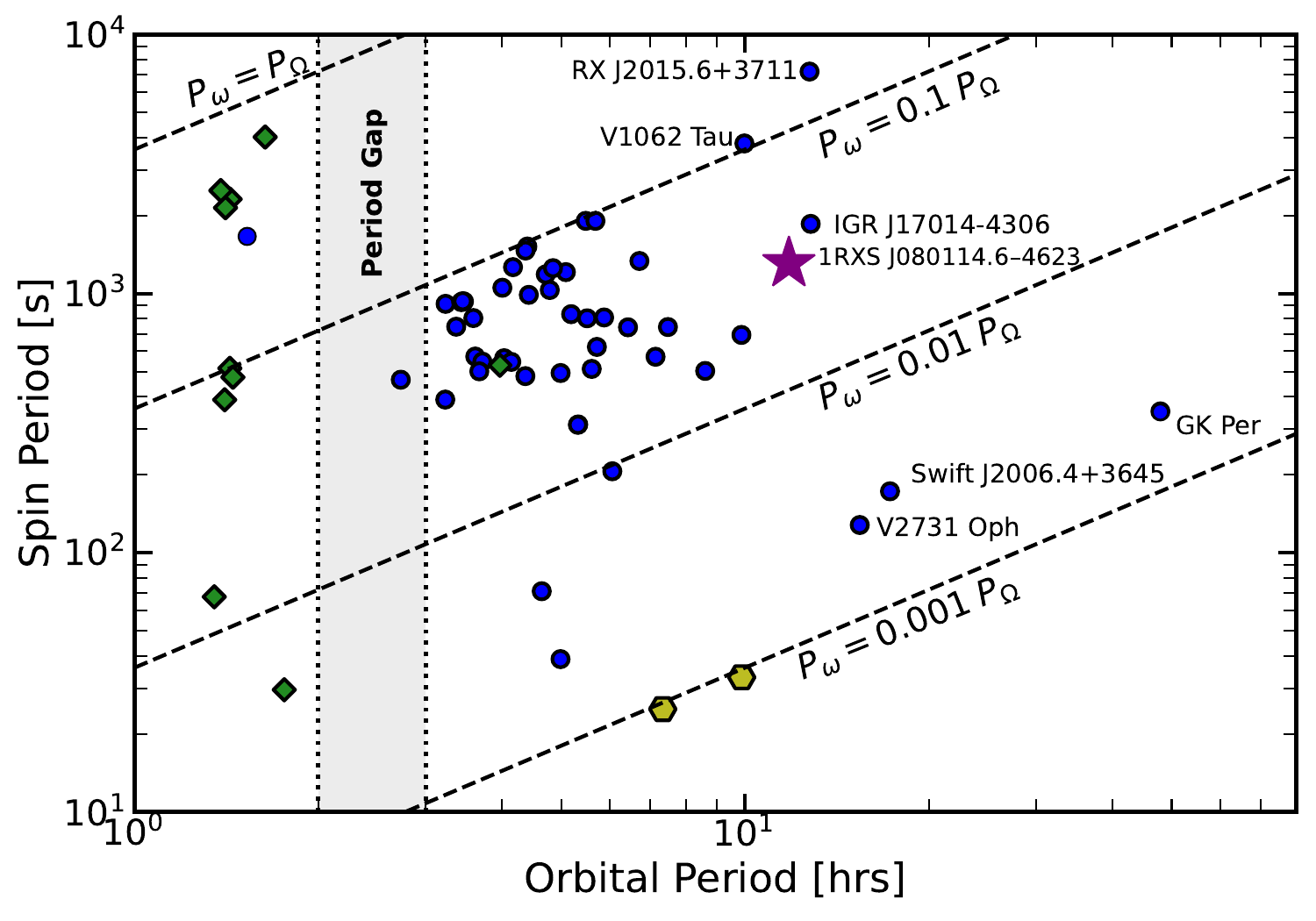}
    \caption{Orbital versus spin period distribution of confirmed IPs, based on data compiled by Koji Mukai. The grey shaded region indicates the CV period gap. The purple star marks the position of 1RXS J080114.6–462324. Green diamonds denote low-luminosity IPs, while the gold hexagon highlights the propeller systems AE Aqr and LAMOST J024048.51+195226.9.}
    \label{fig:Spin_Orbit_Distribution}
\end{figure}

\medskip
\noindent
The identification of the 11.80\,h period as the binary orbital period places 1RXS J080114.6-462324 among the small group of IPs with unusually long orbital periods. Comparable systems include Swift J1701.3-4304 \citep[$P_{\Omega} = 12.8$\,h;][]{2017MNRAS.470.4815B}, RX J2015.6$+$3711 \citep[$P_{\Omega} = 12.7$\,h;][]{2018AJ....155..247H}, and IGR J14091-610 \citep[$P_{\Omega} = 15.84$\,h;][]{2025AJ....169..269J}. Additionally, we constructed a spin-orbital period diagram for all confirmed IPs using the catalogue of Koji Mukai\footnote{The Koji Mukai's catalogue used in the construction of the spin–orbital distribution in this paper is publicly available at: \url{https://asd.gsfc.nasa.gov/Koji.Mukai/iphome/iphome.html}}, including 1RXS J080114.6-462324 and the other known systems (see Fig. \ref{fig:Spin_Orbit_Distribution}). 
The position of 1RXS J080114.6-462324 lies above the period gap. Its spin-to-orbital period ratio is $P_{\omega} / P_{\Omega} \approx 0.0307$, consistent with the characteristics of a disc-fed IP \citep[eg., see][]{2004ApJ...614..349N, 2008ApJ...672..524N}. Assuming spin equilibrium between the WD and accretion disc and using the relation given in Equation (21) of \citet{1994PASP..106..209P}, we estimate the magnetic moment of the WD to be $\mu \approx 2.47 \times 10^{33}\; G\;\rm{cm^{3}}$, which is consistent with a typical magnetic moment for IPs reported by \citet{2004ApJ...614..349N}. Adopting the mass–radius relation of \citet{1972ApJ...175..417N} and considering a typical WD mass range of $0.8$–$0.9 \; M_{\odot}$, we estimate the corresponding WD radius to be $R_{\rm WD} \sim 0.0102$–$ 0.0089 \; R_{\odot}$. 
Using the estimated magnetic moment ($\mu$) and the relation between $\mu$, $R_{\rm WD}$, and the surface magnetic field \citep[see][]{1989MNRAS.237..715N}, we infer a magnetic field strength in the range $B \sim 6.9$–$10.0$ MG, which is typical for IPs (see Section \ref{sec:review}). 
This is comparable with some IPs with inferred magnetic field strength, such as CXOGBS J174517.0-321356 ($ B\gtrsim 7\; \rm MG$; \citealt{2023ApJ...954..138V}) and BG CMi ($ B \approx 5-10\; \rm MG$; \citealt{1987ApJ...322L..35W}).

\medskip
\noindent
Furthermore, as noted in Section \ref{sec:TESS_phot}, the TESS SAP photometric light curve of 1RXS J080114.6–462324, reported by \citet{2024MNRAS.530.3974I}, exhibits a sudden increase in brightness. As expected a similar feature is also present in the TESS PDCSAP light curve. This event was interpreted by the authors as an outburst, possibly a micronova, with supporting evidence provided by simultaneous observations from the All-Sky Automated Survey for Supernovae (ASAS-SN) \citep[see][for more information]{2024MNRAS.530.3974I}.

\subsection{Spectroscopy}\label{sec:spectroscopy_discussion}

The spectrum of 1RXS J080114.6–462324 is typical of that observed in IPs (see Fig. \ref{fig:average_spectrum} and Fig. \ref{fig:H_alpha}), characterised by prominent hydrogen Balmer emission lines such as H$\gamma$, H$\beta$, and H$\alpha$, all of which are associated with accretion processes. Additionally, the presence of He\textsc{ii} $\lambda 4686$ emission provides further indication of a magnetic WD \citep[see, e.g.,][]{1995cvs..book.....W, 1995ASPC...85....3W, 2006A&A...459...21M,2010A&A...519A..96M}. The narrowness of the spectral lines and the absence of any significant modulation in the trailed spectra at the orbital period suggest that 1RXS J080114.6–462324 has a low inclination of $\lesssim$30$^{\circ}$. This is comparable to V2400 Oph, which exhibited a similar spectral profile \citep[see][]{1995MNRAS.275.1028B}. A more detailed inspection of the spectrum reveals red-shifted absorption dips adjacent to H$\beta$ and He\textsc{ii} $\lambda 4686$, observed consistently over six nights of observations (29 March – 3 April 2017). These absorption dips were slightly more pronounced on 3 April 2017. This effect is also evident in the trailed spectra shown in Fig. \ref{fig:trailed_spectra}, where the absorption dips appear to exhibit Doppler variations.

\medskip
\noindent
The power spectrum displayed in the bottom panel of Fig. \ref{fig:Gamma_LS} shows that these absorption features are spin-modulated. We interpret this as the result of infalling material in the accretion curtains obscuring our line of sight to the line-emitting regions near the accretion area, thereby producing spin-phase–dependent absorption features in the observed spectra. Because the system is viewed at a low inclination, the accretion curtain’s infalling material is always moving away from us, producing a persistent redshift in the absorption features we observe.

\medskip
\noindent
It is noteworthy that our recent spectroscopic observations of 1RXS J080114.6–462324 reveal no evidence of a redshifted absorption component adjacent to the H$\alpha$ emission line (see Fig.\ref{fig:H_alpha}). A similar absence is noted for both He\textsc{ii} and H$\beta$ (see bottom panel of Fig.\ref{fig:average_spectrum}), although the signal-to-noise ratio of this particular spectrum is relatively low. This may suggest that the system was in a different accretion state compared to the conditions reported in the 2017 observations.

\subsection{Polarimetry}
\subsubsection{Photopolarimetry}\label{sec:photopol_discussion}

Polarimetric observations of 1RXS J080114.6-462324, obtained with HIPPO, reveal clear detections of circular polarisation (see Fig. \ref{fig:Photopolarimetric}). These results provide direct evidence that 1RXS J080114.6-462324 is a magnetic CV, possessing a relatively strong magnetic field for an IP. The circular polarisation shows peak emission around $\sim+5\%$. The detection of exclusively positive circular polarisation is consistent with the system being of low inclination, limiting our line-of-sight to one accreting pole only.  Fig. \ref{fig:PHOTO_POL_SCARGLE} presents the power spectrum of the circular polarisation percentage flux showing a dominant peak at the WD spin frequency of 66.08 cycles d$^{-1}$. Folding the circular polarisation light curve on the WD spin period further confirms this modulation. The presence of circular polarisation is consistent with cyclotron emission originating from accretion regions channelled by the WD's magnetic field. A few examples of polarised intermediate polars include V2400 Oph \citep{1995MNRAS.275.1028B}, V2731 Oph \citep{2009A&A...496..891B}, NY Lup \citep{2010ApJ...724..165K,2012MNRAS.420.2596P}, and IGR J17014–4306 \citep{2018MNRAS.473.4692P}.

\subsubsection{Circular spectropolarimetry}\label{sec:spectropol_discussion}

The circular spectropolarimetry observations of 1RXS J080114.6–462324 shows that this system is emitting circularly polarised emission at a level of about $+4\%$ over the spectral range 4062–7155\,\AA, centered on the spin phases $\phi = 0.623$ (see Fig. \ref{fig:Spectropolarimetric}). This appears to be consistent with the HIPPO photo-polarimetry (see Fig. \ref{fig: Phased_LC}). The degree of circular polarisation appears to be relatively uniform across this spectral range, with no indication of Zeeman splitting features or any cyclotron humps, which sets a limit to the magnetic strength of the WD to be $\leq$10 MG. This upper bound is consistent with the estimated surface magnetic field of $B \sim 6.9$–$10.0$ MG for a typical IP with a mass range of $0.8$–$0.9 \; M_{\odot}$, and a magnetic moment of $\mu \approx 2.47 \times 10^{33}\; G\;\rm{cm^{3}}$ (see Section \ref{sec:photometry_discussion}).

\section{Conclusion}

In summary, our spectroscopic analysis has revealed unusual spin-modulated redshifted absorption features in the He\textsc{ii} and H$\beta$ emission components. These components resembling inverse P-Cygni profiles indicate inflowing material whose visibility is modulated on the WD's spin frequency of 66.08 cycles d$^{-1}$, revealing a complex accretion geometry which we have attributed to the absorption by the accretion curtain. These findings motivate future high-resolution and time-resolved spectroscopic studies of 1RXS J080114.6–462324 to further constrain the accretion dynamics in this system. We confirm the binary orbital period of $11.8026 \pm 0.0004$ h, as recently reported by \citet{2025ApJS..279...48B}. In addition, we confirm the spin period of the WD to be $1307.5179 \pm 0.0527$ s, in agreement with previously reported values \citep{2017MNRAS.470.4815B,2018AJ....155..247H,2024MNRAS.530.3974I}. We also detect circular polarisation, reaching amplitudes of up to $\sim+5\%$ modulated at the frequency of $66.08$ cycles d$^{-1}$. Independent SALT circular spectropolarimetry observations confirm the presence of circular polarisation at levels of up to $\sim+4\%$. This detection of polarisation provides further evidence for the magnetic nature of 1RXS J080114.6–462324.
 
\medskip
\noindent
It is interesting to note that another long-period IP, IGR J17014–4306, is almost identical to 1RXS J080114.6–462324 in terms of both spin and orbital periods, as well as the detection of spin-modulated circular polarisation \citep{2018MNRAS.473.4692P}. Since IGR J17014–4306 is associated with the nova remnant Nova Scorpius 1437 A.D., and also exhibits recurrent outbursts \citep{2017Natur.548..558S}, continued long-term monitoring of 1RXS J080114.6–462324 will be particularly valuable to establish the frequency and nature of any outbursts or micro-novae \citep{2024MNRAS.530.3974I}. It is also noteworthy that all the long-period IPs ($\gtrsim$9 h) shown in Fig. \ref{fig:Spin_Orbit_Distribution} exhibit circular polarisation \citep{1992ApJ...401..628S, 2025ApJ...984..152L, 2018AJ....155..247H, 2009A&A...496..891B, 2018MNRAS.473.4692P, 2016MNRAS.456.1913C, 2010ApJ...724..165K, 2012MNRAS.420.2596P}, whereas the majority of IPs are generally unpolarised (see: Koji's catalogue of IPs). 

\medskip
\noindent
Monitoring the spin period over longer timescales to determine whether the system is in spin equilibrium will also be important. When combined with an estimate of the white dwarf’s magnetic moment, this will provide key insights into its evolutionary pathway. Simulations by \citet{2004ApJ...614..349N, 2008ApJ...672..524N} show that high-magnetic-moment IPs with long orbital periods are expected to evolve into polars as predicted by \citet{1984ApJ...285..252C}. Interesting evolutionary constraints are provided by atypical cases, including both the asynchronous short-period binary SDSS J134441.83+204408.3 \citep{2023ApJ...943L..24L}, which has a strong (B = 56 MG) WD magnetic field in comparison with the long-period, weaker-field binary of the current study.
By contrast, long-period IPs with lower magnetic moments may follow three possible evolutionary tracks. They may evolve into EX Hya–like systems below the period gap, with $P_{\omega} / P_{\Omega}$ > 0.1, possibly avoiding synchronisation if their secondary stars have weak magnetic fields. 
They may evolve into low-field polars, which could be difficult to observe. 
They may evolve into conventional polars, where the true magnetic field strength is initially buried by high accretion rates but resurfaces once accretion ceases.

\medskip
\noindent
Recent work has also identified a new class of magnetic systems, the prototype being AR Sco \citep{2016Natur.537..374M}, in which strong spin-modulated linear polarisation has been detected \citep{2017NatAs...1E..29B,2018MNRAS.481.2384P}. In these systems, the polarised emission is thought to arise from synchrotron radiation rather than cyclotron emission, suggesting an alternative emission mechanism. 
Taken together, these findings emphasize that further work is required to advance our understanding of how the various subclasses of magnetic cataclysmic variables are related, whether through different evolutionary stages, intrinsic system parameters, or a combination of both. Continued, detailed long-term studies are therefore crucial to expand statistical coverage across parameter space, to identify potential new subclasses, and to uncover any selection effects or observational biases.

\section*{Acknowledgements}

This paper uses spectroscopic and photometric observations taken using the South African Astronomical Observatory (SAAO) facilities, the SAAO 1.0-m, 1.9-m. The circular spectropolarimetry observations reported in this paper were obtained with the Southern African Large Telescope (SALT), under program 2024-2-SCI-038 (PI: ZN Khangale). 
The financial assistance of the South African Radio Astronomy Observatory (SARAO) towards this research is hereby acknowledged (\url{www.sarao.ac.za}). VM,
SBP and DAHB acknowledge research support from the National Research Foundation. 
This paper also includes data from the Transiting Exoplanet Survey Satellite (TESS), which are publicly accessible via the Mikulski Archive for Space Telescopes (MAST). 



\section*{Data Availability}

The TESS data that was used in this paper is publicly available from (\href{https://mast.stsci.edu/portal/Mashup/Clients/Mast/Portal.html}{MAST}).



\bibliographystyle{mnras}
\bibliography{references} 








\bsp	
\label{lastpage}
\end{document}